\documentclass[twocolumn, twocolappendix]{aastex63}
\usepackage{natbib}
\usepackage{graphicx}
\usepackage{CJK}
\usepackage[version=4]{mhchem}
\usepackage{amsmath}
\usepackage{subfigure}
\usepackage{multirow}
\usepackage{wrapfig}
\usepackage{color,xcolor}

\definecolor{ao}{RGB}{0,103,192}
\definecolor{gunen}{RGB}{174,0,46}

\hypersetup{linkcolor=gunen, citecolor=ao, urlcolor=ao}

\graphicspath{{./}{figures/}}

\received{2020.11.5}
\revised{2021.2.23}
\accepted{2021.2.26}
\published{2021.4.22}

\shorttitle{Compact Groups of Galaxies in SDSS \& LAMOST. II. }
\shortauthors{Zheng \& Shen}

\begin{document}
\begin{CJK*}{UTF8}{gbsn}

\title{Compact Groups of Galaxies in Sloan Digital Sky Survey and LAMOST Spectral Survey. II. Dynamical Properties of Isolated and Embedded Groups}

\correspondingauthor{Shi-Yin Shen}
\email{ssy@shao.ac.cn}

\author[0000-0002-5632-9345]{Yun-Liang Zheng (郑云亮)}
\affiliation{Key Laboratory for Research in Galaxies and Cosmology, Shanghai Astronomical Observatory, Chinese Academy of Sciences, 80 Nandan Road, Shanghai, China, 200030.}
\affiliation{University of the Chinese Academy of Sciences,  No.19A Yuquan Road, Beijing, China, 100049.}

\author[0000-0002-3073-5871]{Shi-Yin Shen (沈世银)}
\affiliation{Key Laboratory for Research in Galaxies and Cosmology, Shanghai Astronomical Observatory, Chinese Academy of Sciences, 80 Nandan Road, Shanghai, China, 200030}
\affiliation{Key Lab for Astrophysics, Shanghai, China, 200034.}

\begin{abstract}
Compact groups (CGs) of galaxies appear to be the densest galaxy systems containing a few luminous galaxies in close proximity to each other, which have a typical size of a few tens kiloparsec in observation. On the other hand, in the modern hierarchical structure formation paradigm, galaxies are assembled and grouped in dark matter halos, which have a typical size of a few hundreds of kiloparsec. Few studies have explored the physical connection between the observation-based CGs and halo model-based galaxy groups to date. In this study, by matching the largest local CG catalog of \citet{2020ApJS..246...12Z} to the halo-based group catalog of \citet{2007ApJ...671..153Y}, we find that the CGs are physically heterogeneous systems and can be mainly separated into two categories, the isolated systems and those embedded in rich groups or clusters. By examining the dynamical features of CGs, we find that the isolated CGs have systematically lower dynamical masses than that of noncompact ones at the same group luminosity, indicating a more evolved stage of isolated CGs. On the other hand, the embedded CGs are mixtures of chance alignments in poor clusters and recent infalling groups (substructures) of rich clusters.
\end{abstract}


\section{Introduction} \label{sec:intro}
According to the hierarchical diagram, over half of the galaxies are clustered into the group systems via gravitational instability. Tidal interaction \citep{1972ApJ...178..623T,1992Natur.360..715B}, ram pressures \citep{1972ApJ...176....1G}, and galaxy harassment \citep{1981ApJ...243...32F,1996Natur.379..613M} are expected to be more frequently happen in these bound systems. Group systems of galaxies are therefore widely used to study the environmental dependence of galaxy evolution (e.g., \citealp{1980ApJ...236..351D}, \citealp{1984ApJ...285..426B}, \citealp{2003MNRAS.346..601G}, \citealp{2012A&A...543A.119C}, \citealp{2015MNRAS.451.3249A}, \citealp{2020arXiv200607535C}).

Among various of the group systems, compact groups of galaxies (hereafter CGs) are special cases that typically contain a few luminous galaxies in close proximity to each other. Unlike the massive systems whose higher velocity dispersion favor more rapid fly-byes, the relatively low velocity dispersion ($\sim$ 250 km s$^{-1}$, \citealp{1992ApJ...399..353H}) of CGs make the occurrence of tidal interactions and mergers  to be more frequent (e.g., \citealp{1994ApJ...427..684M}, \citealp{1995ApJ...444L..61O}). 

Historically, \citet{1982ApJ...255..382H} identified 100 CGs from the Palomar Observatory Sky Survey, introducing a set of photometric-based criteria: $(1)$ richness, $(2)$ isolation, $(3)$ compactness. With the follow-up spectroscopic surveys, redshift information has also been absorbed as a criterion (e.g., \citealp{1992ApJ...399..353H, 2012MNRAS.426..296D, 2015JKAS...48..381S}) and optimized for CG selection \citep{2018A&A...618A.157D,2020ApJS..246...12Z}.  In \citet[hereafter Paper I]{2020ApJS..246...12Z}, we have identified a large spectroscopically confirmed CG sample via the modified Hickson criteria:

\begin{enumerate}
\setlength{\itemsep}{-0.7ex}
\item Richness: $3\leq N (14.00 \le r \le 17.77) \leq 10$
\item Isolation: $ \theta_{n} \geq 3$ $\theta_{G}$ 
\item Compactness: $ \mu \leq 26.0$ mag\,arcsec$^{-2}$
\item Velocity Difference: $|V - V_{\text{med}}| \leq 1000$ km s$^{-1}$ 
\end{enumerate}

\noindent where $N$ is the number of members with galactic-extinction-corrected $r$-band Petrosian magnitude $14.00 \le r \le 17.77$, $\mu$ is the $r$-band effective surface brightness (compactness) averaged over the smallest enclosing circle with angular radius $\theta_{G}$, $\theta_{n}$ is the angular radius of the largest concentric circle that contains no external galaxies, $V$ is the recessional velocity of each member, and $V_{\text{med}}$ is their median. 

Among the above Hickson criteria, the compactness ($\mu \le 26.0$) criterion ensures the typical separation of CG members within a few tens of kiloparsecs and the size of CGs ($\theta_G$) typically being less than 100 kpc. However, in $\Lambda$CDM cosmology, the virial radii of the host dark matter halos of  galaxy groups are much more extended. For example, at redshift $z\sim0$, a group of galaxies with halo mass $M_h \sim 10^{13} M_{\odot}$ has a virial radius out to $R_v\sim 500$ kpc. The very much larger $R_v$ compared to $\theta_G$ implies that CGs might be identified or embedded in larger groups as subsystems. Indeed, previous studies (e.g., \citealp{1994PASP..106..413R, 1995AJ....109.1476P, 1998AJ....116.1573B, 2005ASPC..329...67A, 2005AJ....130..425D, 2011MNRAS.418.1409M, 2015A&A...578A..61D}) found a fair proportion of CGs ($\sim 20 \% - 95 \%$) are coincident within substructures of larger systems.

As a result, the physical nature and the boundness of CGs have been debated for several decades. Both theoretical models (e.g., \citealp{1986ApJ...307..426M, 2006A&A...456..839T}) and simulation studies (e.g., \citealp{1995ApJ...442...57H, 2010MNRAS.409.1227D, 2020MNRAS.492.2588D, 2020MNRAS.491L..66H}) have suggested that CGs are a mixture of virialized groups, chance alignments in filaments, and collapsing groups as bound substructures within clusters. An alternative scenario has also been proposed that the large groups are the birthplace of the embedded CGs \citep{1994AJ....107..868D,1994AJ....107.1623R,2005ASPC..329...67A}. However, the studies on dynamic links between the observationally identified CGs and the dynamical systems defined in larger scales (e.g., normal groups or clusters) are still limited \citep{1998AJ....116.1573B, 2008A&A...486..113M, 2015A&A...578A..61D}.

Many samples of galaxy groups based on halo models have been established by various redshift surveys: e.g., \citet{2005MNRAS.356.1293Y} from the 2dFGRS; \citet{2006MNRAS.366....2W}, \citet{2007ApJ...671..153Y, 2012ApJ...752...41Y}, \citet{ 2012MNRAS.423.1583M}, and \citet{2020A&A...636A..61R} from the SDSS; \citet{2016ApJ...832...39L} from the 2MRS; \citet{2017MNRAS.470.2982L} from the 2MRS, 2dFGRS, 6dFGRS, and SDSS.  Meanwhile, different CG catalogs also have been constructed (e.g., \citealp{1996AJ....112..871B, 2004AJ....127.1811L, 2009MNRAS.395..255M, 2012MNRAS.426..296D, 2016ApJS..225...23S, 2018A&A...618A.157D}). The completely different selection philosophy between the halo-based galaxy groups and Hickson-like CGs make the study of the connection between the two categories be quite complex.  For gravitationally bound systems, the velocity dispersion of their members is a good indicator of their dynamical mass. However, because traditional Hickson-like CGs are selected from photometric-only criteria, their dynamical properties have not been carefully addressed yet. 

In this study, we take advantage of the large spectroscopically selected CG samples in Paper I and aim to further probe their dynamical status through the line-of-sight (LOS) velocity dispersion. We expect to use velocity dispersion as a dynamical indicator to further clarify the physical connections between the Hickson-like CGs and halo-based group systems. The layout of this paper is organized as follows. In section~\ref{sec:data}, we briefly describe the CG samples and investigate their connection to the halo-based groups defined by \citet{2007ApJ...671..153Y}. We further divide the CG sample into different categories based on their different correlations with the halo-based groups. We then present a detailed dynamical analysis of the different categories of CGs and give relevant discussions in section~\ref{sec:results}. Finally, we summarize our results in section~\ref{sec:summarize}.

Throughout this paper, we assume the flat WMAP7 cosmology with parameters $H_{0} = 70$ km s$^{-1}$ Mpc$^{-1}$ and $\Omega_{m} = 0.27$ \citep{2011ApJS..192...18K}.

\section{Samples and Data} \label{sec:data}
\subsection{Compact Group Sample} \label{sec:cg}
In Paper I, CGs were derived from the latest version of New York University Value-Added Galaxy Catalog (VAGC, \citealp{2005AJ....129.2562B}) which is based on the SDSS legacy survey with a set of improved reduction. The redshift incompleteness due to fiber collision has been reduced by SDSS-DR14 \citep{2018ApJS..235...42A}, GAMA-DR2 \citep{2015MNRAS.452.2087L}, and LAMOST-DR7 \citep{2015RAA....15.1095L}. As presented in Paper I, we selected CGs in redshift slices to reduce the bias against nearby groups and derived 6144 conservative CGs (hereafter cCGs) containing 19,465 galaxies with complete redshifts. Also, as mentioned in Appendix A of Paper I, 74 cCGs have an association with bright galaxies ($r < 14.00$ mag). When we join these bright galaxies into the 74 cCGs, 26 of them are still identified as cCGs but their richness and $L_{19.5}$ (see Section~\ref{sec:lumgr}) would be updated, while the other 48 violate the CG criteria and thus would be removed from the CG sample used in this paper. In addition, we further remove 16 CGs after a careful inspection of the spectroscopic data of all group members, where the spectroscopic redshifts of a few member galaxies have been incorrectly reused during the CG construction and therefore could potentially bias the velocity dispersion finally measured. This results in a sample of 6080 cCGs with 19,273 member galaxies.

\subsection{Halo-based Group Sample: Y07 group catalog} \label{sec:y07}
In this paper, we adopt the galaxy group catalog constructed by \citet[hereafter Y07]{2007ApJ...671..153Y, 2012ApJ...752...41Y}, which is also based on the VAGC of SDSS-DR7. Y07 has applied a halo-based group finder to assign each galaxy in the SDSS-DR7 Main Galaxy Sample (MGS) within the redshift range $0.01 < z < 0.20$ to a unique group. In Y07, three versions of the group catalog have been constructed based on different redshift sources. In this work, we use sample III of Y07 group catalog, where a small fraction of SDSS MGS ($\sim 37,000$) without spectroscopic redshifts have been assigned  with the redshifts of their nearest neighbors. Therefore, the sample III of Y07 group catalog has a 100\% completeness of member galaxies, while also containing contamination from background or foreground galaxies. 

Because a significant amount of SDSS MGS ($\sim 13,000$) have achieved new spectroscopic redshifts from several latest surveys \citep{2016RAA....16...43S, 2019ApJ...880..114F}, we use them to update the sample III of Y07 group catalog. Among these galaxies with new spectroscopic redshifts, $\sim 8500$ have concordant redshifts with their nearest neighbors, thus these groups remain unchanged. For the remaining $\sim 4500$ galaxies with newly measured spectroscopic redshifts, we simply disentangle them from their initial groups (mostly $N = 2$) and update their group luminosity $L_{19.5}$ accordingly (see Section~\ref{sec:lumgr}). For the other groups that still have members with assigned redshifts, we keep them unchanged. As we will show in the next section, these assigned redshifts have negligible effects on our study.

\begin{figure*}[t!]
  \centering
  \subfigure{
    \includegraphics[width=.8\columnwidth]{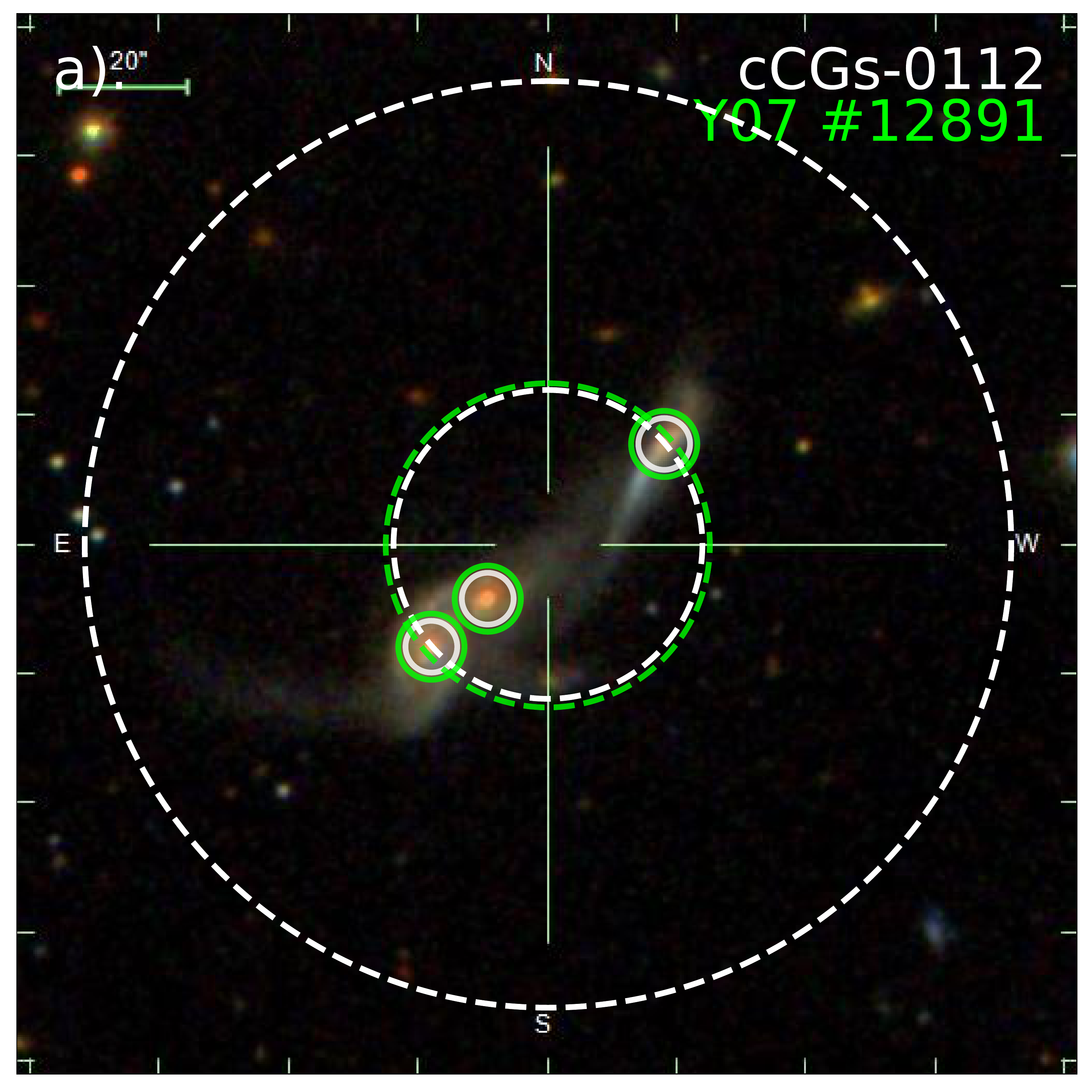}}
  \subfigure{
    \includegraphics[width=.8\columnwidth]{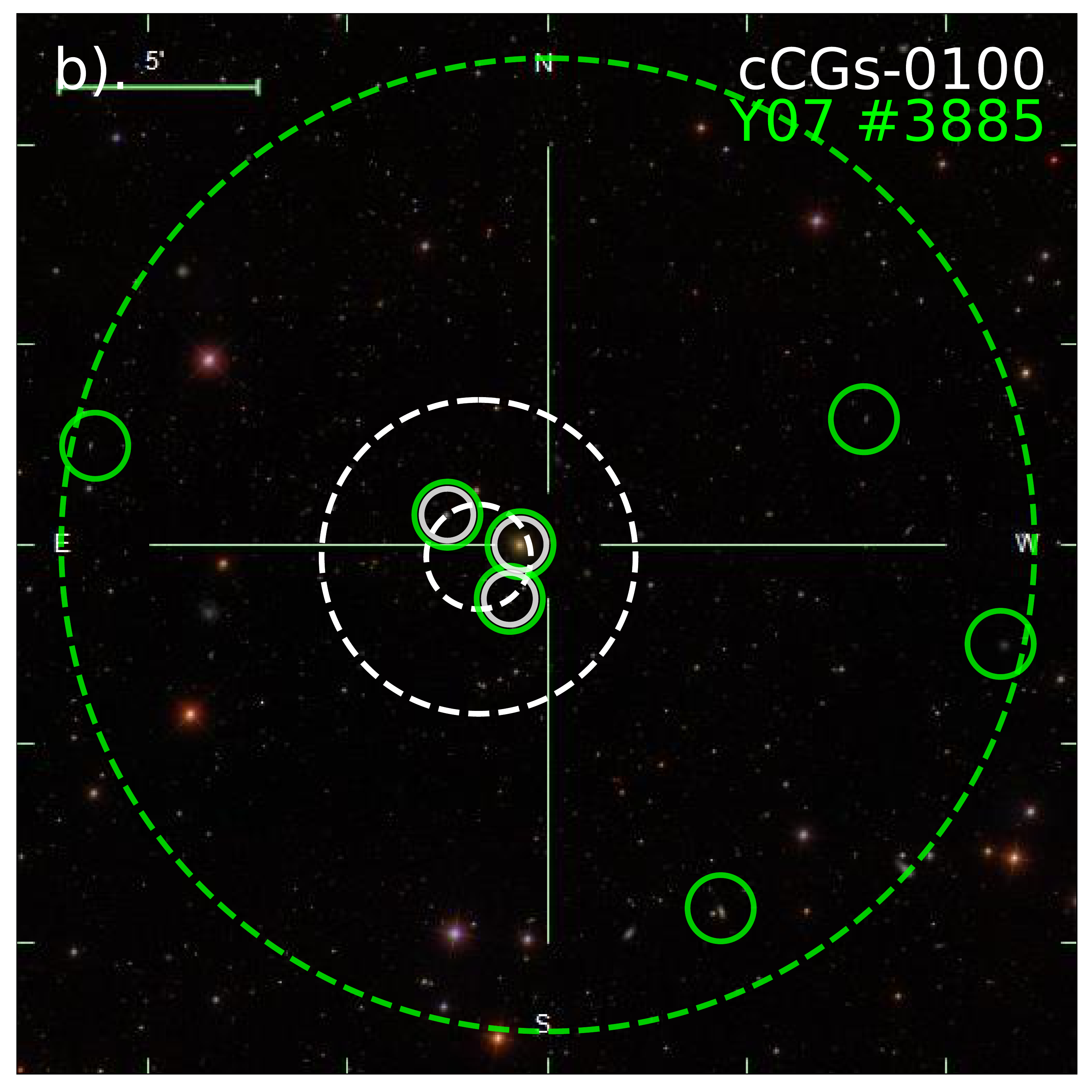}}
  \subfigure{
    \includegraphics[width=.8\columnwidth]{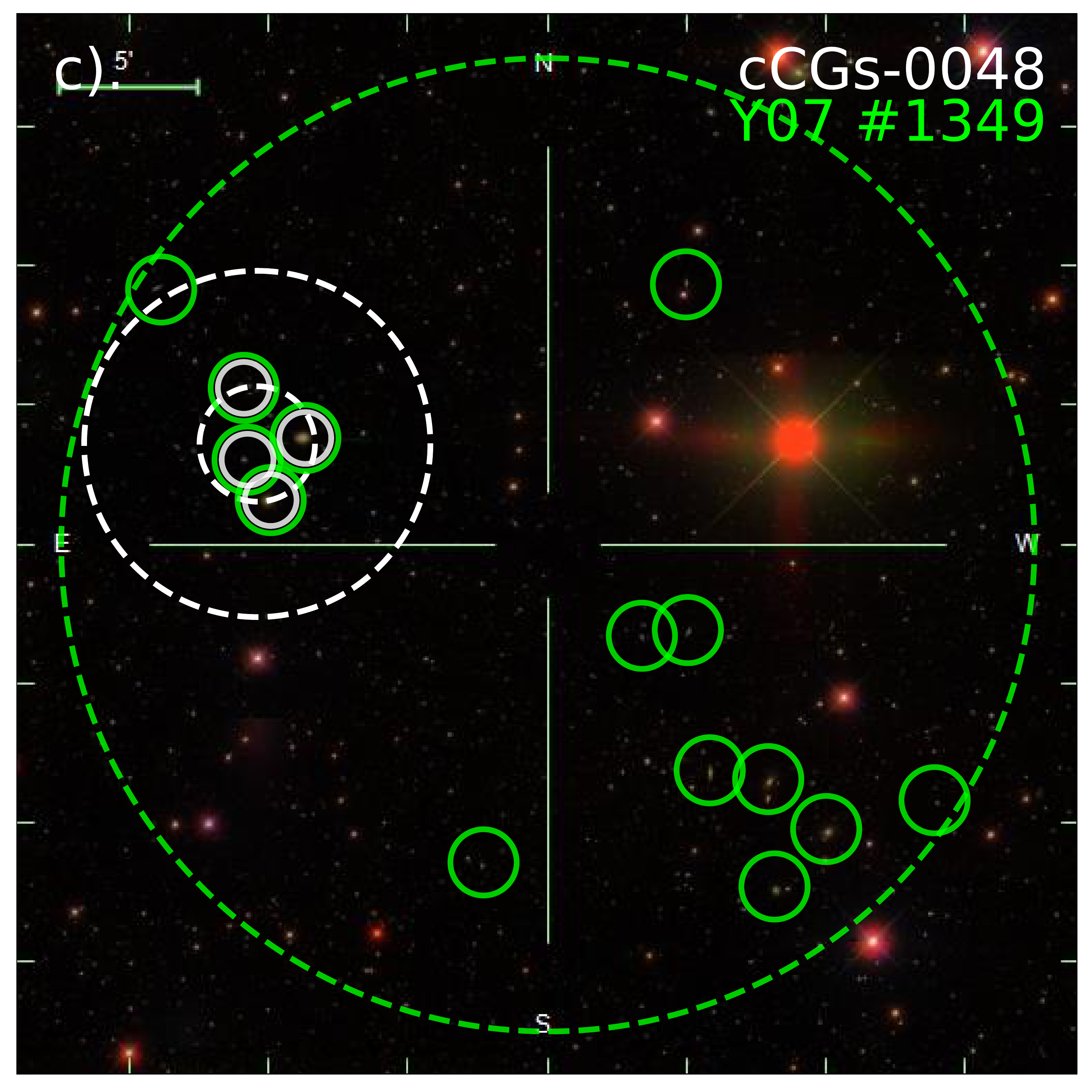}}
  \subfigure{
    \includegraphics[width=.8\columnwidth]{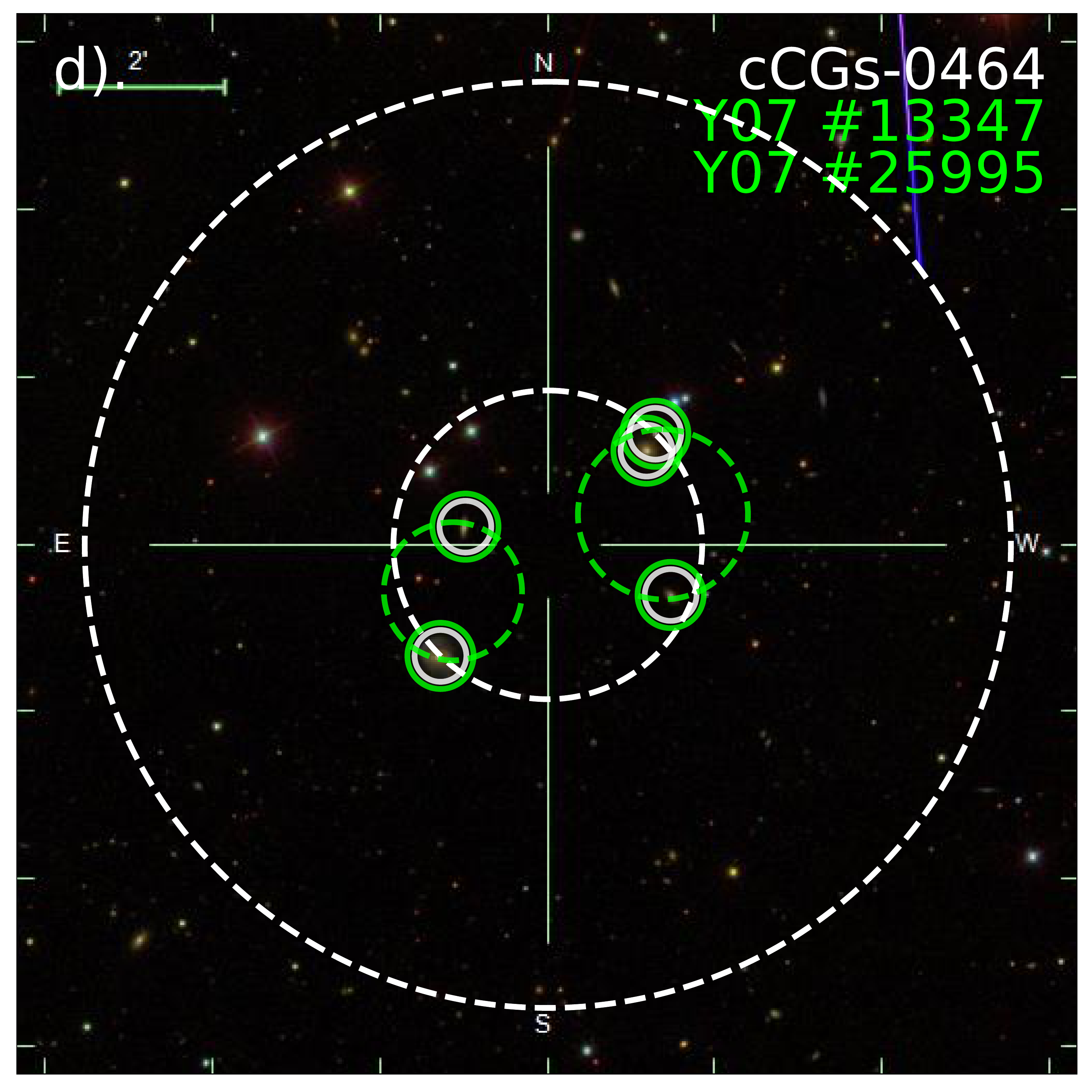}}
  \caption{Example SDSS images of cCGs where their members are in one-to-one correspondence with the members of Y07 groups: (a) isolated CGs with no external galaxy host in the same halo. (b) Predominant CGs with other fainter galaxies sharing the same halo.  (c) Embedded CGs with brighter galaxies occupying the same halo as nondominant subsystems. (d) Split CGs whose members belong to at least two different halos of Y07.  The inner white dashed circles represent the smallest enclosed circles $\theta_G$, the outer white dashed circles represent the concentric circles 3$\theta_G$. Green dashed circles represent the smallest enclosed circles for the Y07 groups, which are manually enlarged for clarity. Solid circles mark the member galaxies of cCGs (white) or their corresponding Y07 groups (green). The ID of the cCG and its corresponding Y07 group are labeled at the top-right corner of each image. }
  \label{fig:CaseN}
\end{figure*}

\begin{deluxetable}{lc}
\tablecaption{The classification of CG subsamples \label{tab:classification}}
\tablewidth{0pt}
\tablehead{
\colhead{CG Subsamples \tablenotemark{}} &
\colhead{Sample Size \tablenotemark{}}} 
\startdata
Isolated CGs & 1667 \\
Predominant CGs & 1570 \\
Embedded CGs: & 1370 \\
\quad \footnotesize Single Embedded Group & \footnotesize 901 \\
\quad \footnotesize Multiple Embedded Groups & \footnotesize 469 \\
Split CGs & 1282 \\
Unmatched CGs & 191 \\
\hline
Overall & 6080 
\enddata
\end{deluxetable}

\begin{figure*}
  \centering
  \subfigure{
    \includegraphics[width=.8\columnwidth]{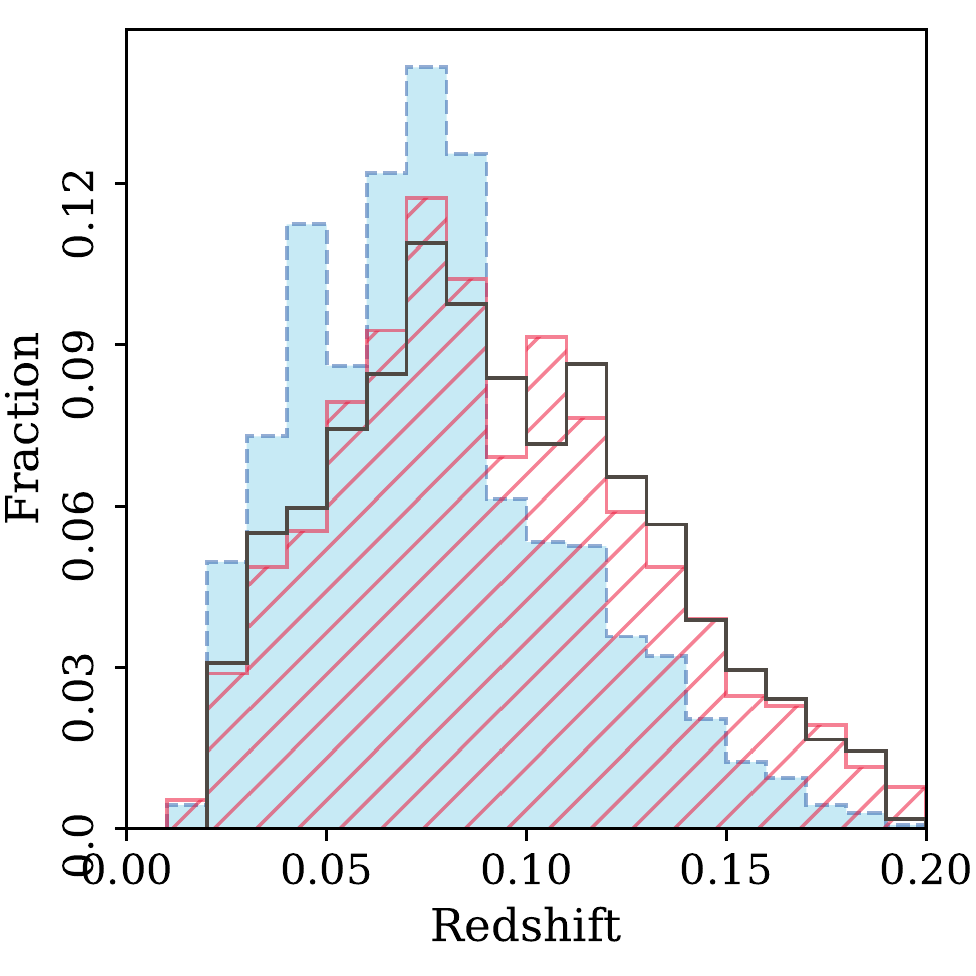}}
  \subfigure{
    \includegraphics[width=.8\columnwidth]{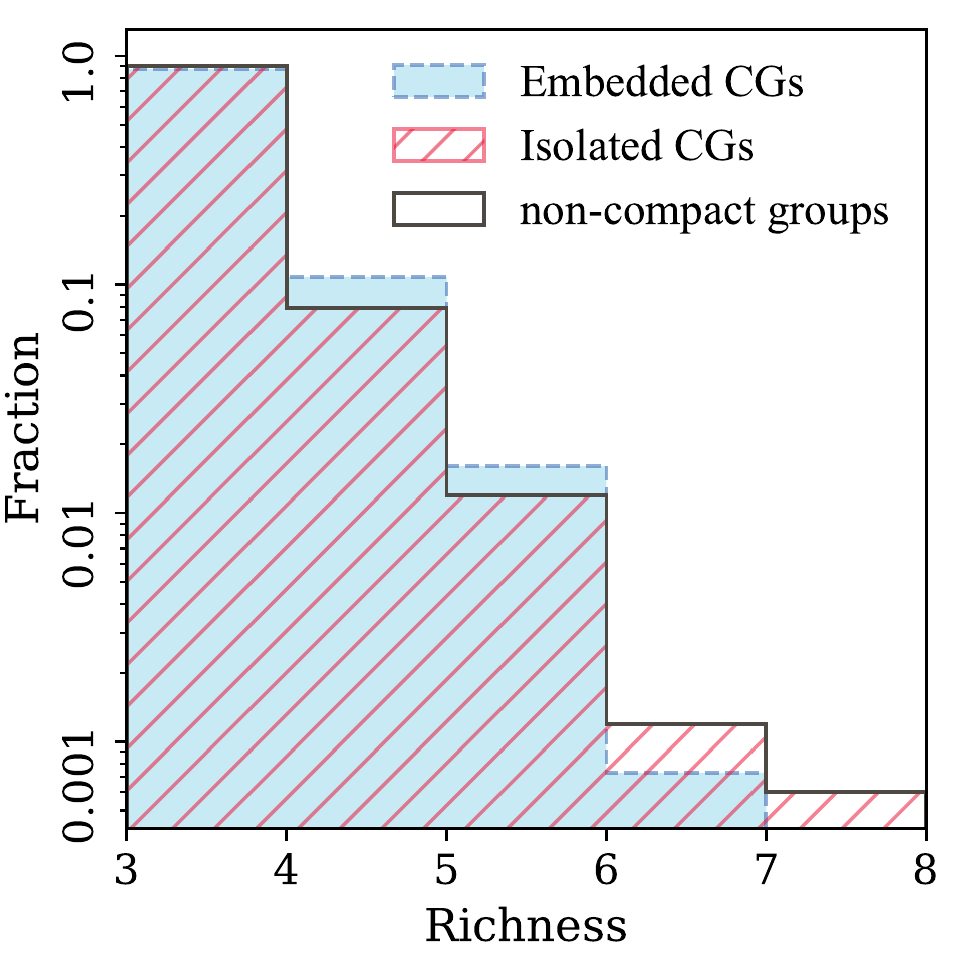}}
  \subfigure{
    \includegraphics[width=.8\columnwidth]{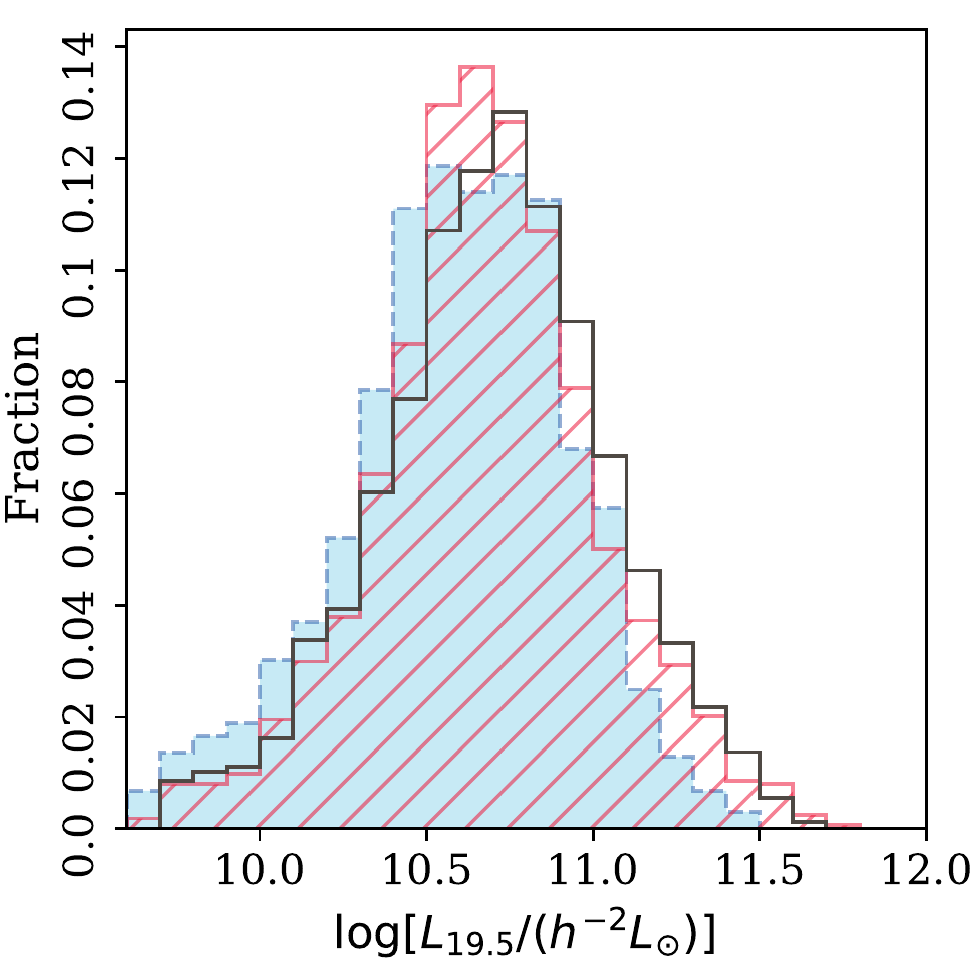}}
  \subfigure{
    \includegraphics[width=.8\columnwidth]{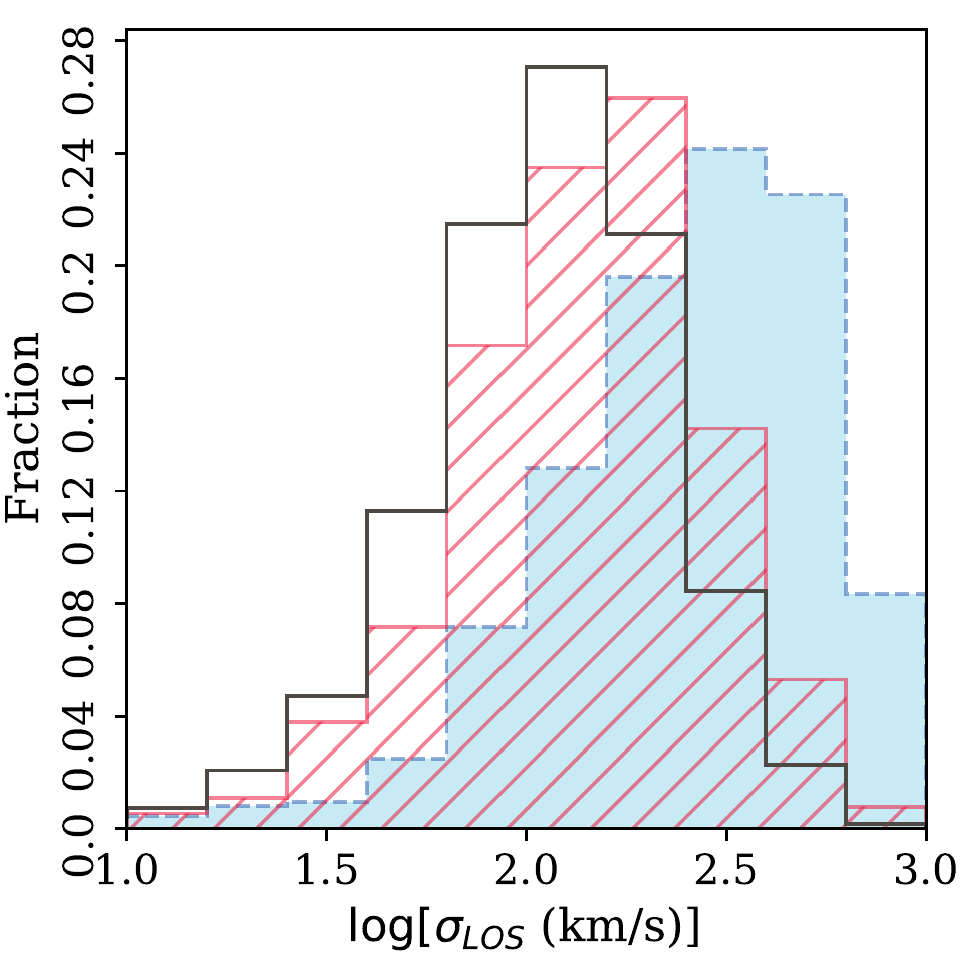}}
  \caption{The distributions of the isolated (red hatched) CGs, embedded (blue filled) CGs, and control sample of noncompact groups (grey open). Upper left: redshift. Upper right: richness. Lower left: group luminosity, $L_{\text{19.5}}$. Lower right: LOS velocity dispersion, $\sigma_{\text{LOS}}$.}
  \label{fig:Basic}
\end{figure*}

\subsection{CG Categories: match with Y07 groups} \label{sec:iso/emb}
We cross-match the members in cCGs with the updated sample III group catalog shown above. Most of the galaxies ($\sim 94 \%$) in cCGs have their group identity in Y07, while the remaining $\sim 6 \%$ galaxies have no corresponding Y07 groups due to one of the two following reasons:

\begin{enumerate}
\setlength{\itemsep}{-0.7ex}
\item The sky coverage of the galaxy catalog we used is slightly larger than that of Y07, because Y07 has discarded the galaxies located near the survey edge or in very low completeness regions. This results in 118 cCGs have no counterparts in Y07.
\item In Y07, the faint-end magnitude cuts vary with position, ranging from 17.62 to 17.72 in extinction-corrected Petrosian magnitude, which are the results of different versions of target selection of the MGS for spectroscopic observation through the period covered by the Early Data Release \citep{2002AJ....123..485S}, while we adopt the latest version of a fixed value $r_{f} = 17.77$ mag for CG selection. This operation results in the updating of $\sim 470$ Y07 groups. We have tested that almost all of the extra members indeed belong to the same Y07 groups according to the Y07 group finder. However, there are also 73 cCGs with at least 2 members without a match in Y07 groups. For safety, we discard them from further investigation.
\end{enumerate}

As a result, 5889 out of 6080 cCGs with all of their members could be matched with the updated Y07 groups. Among them, 1667 have the same members as the Y07 groups, while 2940 cCGs are the subsets of Y07 groups. Apart from these one-to-one matches, there are also 1282 cCGs with their members matched to different Y07 groups. These split CGs are mainly attributed to the large velocity difference cut $\Delta V < 1000$ km s$^{-1}$ used in our CG selection, while the velocity difference cut used in the Y07 groups is dynamically linked to their virial mass. 

In this study, we ignore the split CGs (see Appendix~\ref{sec:multi} for a detailed discussion). We define these 1667 CGs with the same memberships as the Y07 groups as isolated CGs.  For the cCGs that are subsets of Y07 groups, we define them as ``embedded systems," where their host Y07 groups are referred to as ``parent groups." To better distinguish the dynamical effects induced by parent groups, we define the embedded CGs as the subgroups that do not dominate the luminosity of their parent groups according to
\begin{equation}
\sum^{N_{\text{par}}}_{i=1} L_{i} \ge 2\sum^{N_{\text{emb}}}_{j=1} L_{j}, \label{eq:emb}
\end{equation}
where $L_{j}$ is the luminosity of the $j{\rm th}$ member of embedded CGs and $L_{i}$ is the luminosity of the $i{\rm th}$ member of their parent groups. For each galaxy, we compute their $^{0.1}r$-band luminosity using
\begin{equation}
\frac{L}{L_{\odot}} = 10^{-0.4\left[r-\text{DM}(z)-K^{0.1}_r(z)-4.64\right]},
\end{equation}
where $\text{DM}(z)$ is the bolometric distance modulus, $K^{0.1}_r(z)$ is the K-correction value at $z = 0.1$ calculated using the \texttt{KCORRECT} package of \citet{2007AJ....133..734B}, and $4.64$ is the $r$-band magnitude of the sun in AB system. 

This leaves 1370 embedded CGs being hosted by 1084 parent groups, while the 1570 remainders are referred to as predominant CGs. We show example images for each type of CG in figure~\ref{fig:CaseN}. Table~\ref{tab:classification} summarizes the results of the classification. In the following, we do not consider the predominant CGs because it is difficult to distinguish the dynamical effects between the embedded system and host groups. A basic comparison of the parent groups of predominant CGs and embedded CGs is presented in Appendix~\ref{sec:parg}.

For embedded CGs, there are cases where multiple CGs exist in a single host galaxy group. In our sample, there are 469 CGs in this situation, hosted by a total of 183 parent groups. We have tested that these multiple embedded CGs and single embedded CGs did not differ statistically in their dynamical properties (see Appendix~\ref{sec:parg} for details). Therefore, for statistical significance, we do not distinguish between these two cases in the later sections and uniformly refer to them as embedded CGs.

It is worth mentioning that $\sim 45 \%$ of the parent groups have few member galaxies with redshifts assigned from their nearest neighbors (Section~\ref{sec:y07}). These redshift-assigned members would be ignored when we calculate the velocity dispersion of the parent groups. On the other hand, these members are included in calculating their total luminosity (Section~\ref{sec:lumgr}). As most of the parent groups of embedded CGs contain $N \gtrsim 8$ galaxies, the including or excluding of these redshift-assigned members have negligible effects on the results of this study.

Besides the above two categories of CGs, we also built a control sample of noncompact groups of galaxies with compactness $\mu > 26.0$ mag arcsec$^{-2}$ from the Y07 group catalog. To do that, we first exclude the groups with incomplete spectroscopic redshifts and containing any galaxies in cCGs. Then, we match their richness and redshift distribution to the isolated CG sample. For each isolated CG,  we match three unique Y07 groups which have the same richness and redshift within a tolerance of $z \sim 0.01$ by means of a Monte Carlo sampling. Finally, we get a control sample of $3 \times 1667$ ``noncompact groups of galaxies."

We show the histograms of redshift and richness for the isolated CGs, embedded CGs, and the control noncompact groups in the upper-left and right panels of figure~\ref{fig:Basic}. Actually, the richness distributions of the isolated and embedded CGs are quite similar, while the redshift distribution of embedded CGs is biased to lower redshifts. This bias in redshift is mainly caused by the fact that high-richness groups (parent groups of embedded CGs) are biased to low redshifts in a flux-limited sample.

\begin{figure}
  \centering
  \subfigure{
    \includegraphics[width=1.\columnwidth]{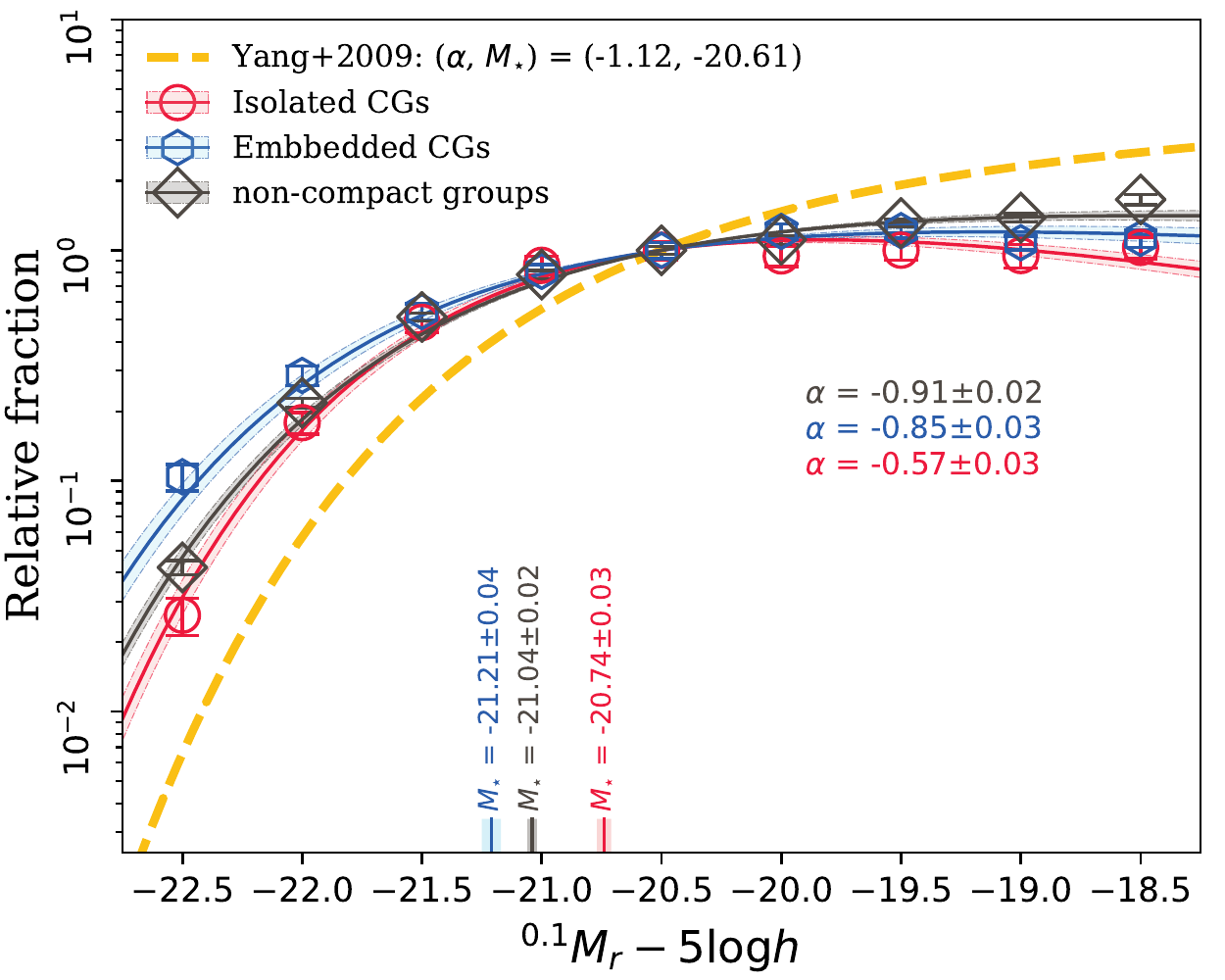}}
  \caption{The $^{0.1}r$-band luminosity functions of isolated (red), embedded (blue) CGs, and noncompact groups (black) derived via both a nonparametric (stepwise) and parametric maximum likelihood estimator with ($\alpha$, $M_{\star}$) quoted inside the figure. The dashed line represents the LFs of all the galaxies in Y07 catalog given by \citet{2009ApJ...695..900Y}. All of the LFs are normalized to 1.0 at $^{0.1}M_{r} - 5\log{h} = -20.5$ mag for comparison.}
  \label{fig:LFs}
\end{figure}

\subsection{Total Group Luminosity: $L_{19.5}$} \label{sec:lumgr}
For an unbiased comparison of galaxy groups at different redshifts, a characteristic total luminosity of galaxy groups needs to be defined. In Paper I, we simply summed up the luminosities of the members for all CGs and argued that this `apparent total group luminosity' is a good proxy of their real total luminosity. In this study, following Y07, we use $L_{19.5}$ to characterize the total luminosity of each galaxy group, which is defined as the sum of the luminosities of all members brighter than $^{0.1}M_{r} - 5\log{h} < -19.5$ mag. For groups with $z \le 0.09$, the faint-end flux limit $r_f \sim 17.77$ mag of the SDSS MGS ensures that all their members with $^{0.1}M_{r} - 5\log{h} < -19.5$ mag had been targeted. Therefore, the $L_{19.5}$ of these groups are obtained by summing up the luminosities of the group members brighter than $^{0.1}M_{r} - 5\log{h} < -19.5$ mag directly. For distant ($z > 0.09$) groups, we make a correction to the observed total luminosity using
\begin{equation}
L_{19.5} = \frac{\int^{\infty}_{L_{\text{cut}}}L\Phi\left(L\right)dL}{\int^{\infty}_{L_{\text{f}}(z)}L\Phi\left(L\right)dL} \sum^{N}_{i=1} L_{i}, \label{eq:L195}
\end{equation}
\noindent where $L_{\text{cut}}$ is the luminosity that corresponds to $^{0.1}M_{r} - 5\log{h} = -19.5$ mag, $L_{\text{f}}(z)$ is the faint luminosity limit of a galaxy that can be observed at the redshift of that group, and $\Phi\left(L\right)$ is the $^{0.1}r$-band luminosity function (hereafter LF) of the group members being considered. We use two canonical methods to derive the LFs of these samples: the nonparametric stepwise maximum likelihood \citep{1988MNRAS.232..431E} for binned LF and the maximum likelihood estimator \citep{1979ApJ...234..775T} to calculate the best-fit \citet{1976ApJ...203..297S} function:
\begin{equation}
{\Phi\left(L\right)}dL = {\Phi}_{\star} {\left(\frac{L}{L_{\star}}\right)}^{(\alpha+1)} \exp{\left(-\frac{L}{L_{\star}}\right)} d\left(\frac{L}{L_{\star}}\right), \label{eq:LF}
\end{equation}
\noindent where ${\Phi}_{\star}$ is the overall amplitude, $L_{\star}$ is the characteristic luminosity, and $\alpha$ is the faint-end slope. 

We calculate the LFs of the members for the isolated, embedded CGs, and noncompact groups, respectively. We plot them in figure~\ref{fig:LFs} where the LF of the all Y07 galaxies by \citet{2009ApJ...695..900Y} is also shown for comparison. Here we focus only on the shape of these LFs, and all the LFs are normalized to 1.0 at $^{0.1}M_{r} - 5\log{h} = -20.5$ mag for comparison. As can be seen, firstly, the two different LF estimators give consistent results. Next, the LFs of the galaxies in groups show systematical differences from those of all the galaxies, which have a brighter characteristic magnitude and shallower faint-end slope. This systematical difference might be the results of the conditional LFs of the galaxies in different halos having a systematical mass dependence \citep[e.g.,][]{2003MNRAS.339.1057Y, 2003MNRAS.340..771V, 2011MNRAS.415.2553Z}. 

On the other hand, the isolated CGs and noncompact groups, although having the same richness distribution, also show significant differences in their LFs, which implies that the compactness of groups has a nurturing effect on the evolution of their member galaxies. Such a difference might also be attributed to the different dynamical states of the group systems (e.g, \citealt{2012MNRAS.419L..24M}). Moreover, the LFs of embedded CGs also show differences from those of isolated CGs, which may have an even more complex physical origin because the richness of the parent groups of the embedded CGs is significantly larger than that of the isolated ones.

However, we will not explore the physical implications of the LFs of CGs in more detail in this study, but simply use them in Equation~\ref{eq:L195} for total luminosity calculation. A more detailed study of the physical properties of the CG member galaxies will be presented in an upcoming study.

With LF calculated using Equation~\ref{eq:L195},\footnote{Note that there are 77 CGs without $L_{19.5}$ calculated because that none of their members is brighter than $^{0.1}M_{r} - 5\log{h} < -19.5$ mag.}, we show the distributions of the final  $L_{19.5}$ of the isolated CGs, emebeed CGs, and control noncompact groups in the lower-left panel of figure~\ref{fig:Basic}. As a result of the lower redshift distribution (upper-left panel of figure~\ref{fig:Basic}), the embedded CGs show a systematically lower $L_{19.5}$ distribution than the isolated ones.

\subsection{Velocity Dispersion} \label{sec:sigv}
The LOS velocity dispersions of the groups are computed using a variant of the gapper estimator described by \citet{1990AJ....100...32B}, which is less biased for small groups \citep{2015A&A...578A..61D}. The method involves ordering the set of recessional velocities $\left\{V_i\right\}$ of the $N$ member galaxies and defining gaps as
\begin{equation}
g_i = V_{i+1} - V_{i},  \qquad i = 1, 2, \cdots, N-1
 \label{eq:gapper}
\end{equation}
The rest-frame LOS velocity dispersion is given by \citet{2005MNRAS.356.1293Y}:
\begin{equation}
\sigma_{\text{gap}} = \frac{\sqrt{\pi}}{\left(1 + z_{g}\right) N(N-1)} \sum_{i=1}^{N-1} w_{i} g_{i}
 \label{eq:sigma}
\end{equation}
\noindent where $z_{g}$ is the group redshift and $w_{i}$ is the Gaussian weight defined as $w_{i} = i(N-i)$. 

In practice, if we assume one of the members is static at the center-of-mass velocity of that group, the estimated $\sigma_{\text{gap}}$ therefore should be multiplied by an extra factor $\sqrt{N/(N-1)}$ following \citet{2004MNRAS.348..866E}. Also, the redshift measurement errors increase the estimate of $\sigma_{\text{LOS}}$ in quadrature, thus the final $\sigma_{\text{LOS}}$ of each group is given by
\begin{equation}
\begin{aligned}
\sigma_{\text{LOS}} &= \sqrt{\text{max}\left(0, \frac{N \sigma_{\text{gap}}^{2}}{N-1}-V_{\text{err}}^{2}\right)} \\
V_{\text{err}}^{2} &= \frac{1}{N}\sum^{N}_{i=1}V_{\text{err},i}^{2}  \label{eq:los}
\end{aligned}
\end{equation}
\noindent where $V_{\text{err},i}$ is the recessional velocity error of the $i{\rm th}$ member of the group. In most of the cases, the contribution from $V_{\text{err},i}$ is negligible. The typical value of $V_{\text{err}}$ of the galaxies in the SDSS and LAMOST Spectral Survey is at the level of $\sim 10$ km s$^{-1}$. For the redshifts taken from alternative surveys (e.g., 2dFGRS and GAMA) and without errors for individual galaxies, we use the typical uncertainty $V_{\text{err},i} \sim 33$ km s$^{-1}$ for GAMA \citep{2014MNRAS.441.2440B} and $\sim 60 - 120$ km s$^{-1}$ for 2dFGRS (depends on spectroscopic quality; \citealt{2001MNRAS.328.1039C}).

There is significant randomness in calculating the $\sigma_{\text{LOS}}$ of small groups, which results in and dominates the error of $\sigma_{\text{LOS}}$ of each CG. We denote the error of  $\sigma_{\text{LOS}}$ by $\sigma_{\text{err}}$. We estimate the $\sigma_{\text{err}}$ of each CG by performing a simple Monte Carlo simulation. More specifically, for a  group with $N$ members and estimated  $\sigma_{\text{LOS}}$,  we randomly generate $\sim$ 100,000 sets of mock groups with $N$ recessional velocities from the Gaussian distribution $\mathcal{N}\left(0,{\sigma}_{LOS}^{2}\right)$. We then calculate the velocity dispersion for each mock group using Equation~\ref{eq:los} and take the scatter of 100,000 mock groups as the expected value of $\sigma_{\text{err}}$. 

\begin{figure}
  \centering
  \subfigure{
    \includegraphics[width=1.\columnwidth]{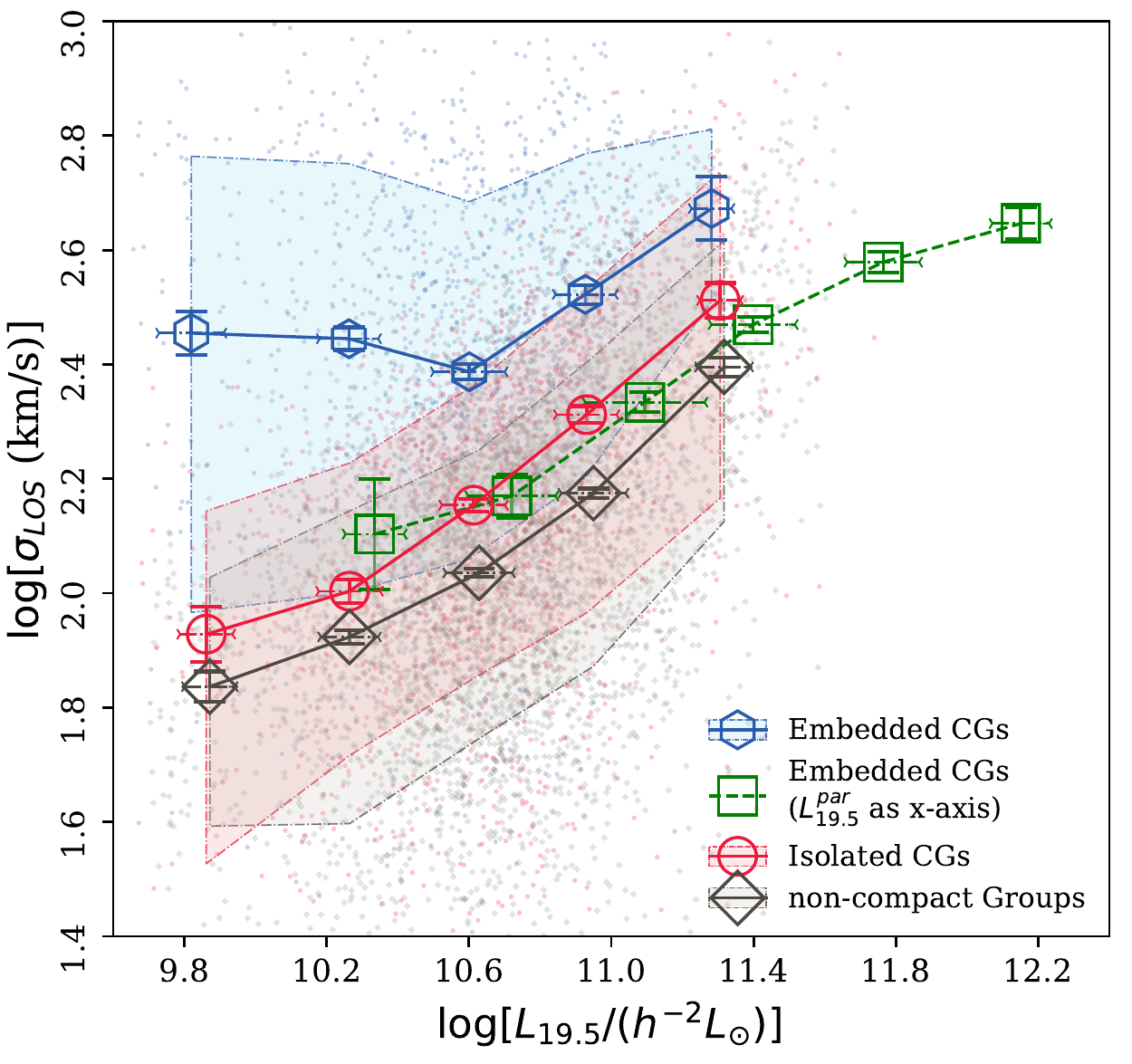}}
  \caption{Group velocity dispersion ($\sigma_{\text{LOS}}$) as a function of group luminosity ($L_{19.5}$) for isolated CGs (red), embedded CGs (blue), and control noncompact groups (black) on a logarithmic scale with bin size of $0.4$ dex. The open circles, hexagons, and diamonds show the median of $\sigma_{\text{LOS}}$ in each $L_{19.5}$ bin, whereas the 16$^\text{th}$ and 84$^\text{th}$ percentiles are covered by shaded areas.  Only the data bins with at least 10 groups are plotted. The green line represents the same scale relation for embedded CGs but their $L_{19.5}$ are replaced by $L_{19.5}^{\text{par}}$. The vertical error bars show the errors of the median $\log\left(\sigma_{\text{LOS}}\right)$ and the horizontal error bars indicate the median absolute deviation of $\log\left(L_{19.5}\right)$ in each bin}.
  \label{fig:L195Dyn}
\end{figure}

We show the distributions of $\sigma_{\text{LOS}}$ of the isolated CGs, embedded CGs, and the control noncompact groups in the lower-right panel of figure~\ref{fig:Basic}. Although the embedded CGs have a similar richness distribution (upper-left panel of figure~\ref{fig:Basic}) and an even lower redshift distribution (upper-right panel of figure~\ref{fig:Basic}) than isolated CGs, they have a systematically higher $\sigma_{\text{LOS}}$ distribution than isolated ones. This systematical difference implies a different physical origins of these two types of CGs, which we will discuss next.

\section{Results} \label{sec:results}
\subsection{$\sigma_{\text{LOS}} - L_{19.5}$ Relation} \label{sec:Lsig}
Figure~\ref{fig:L195Dyn} displays the scale relations between the median $\sigma_{\text{LOS}}$ and $L_{19.5}$ for isolated CGs, embedded CGs, and noncompact groups, where each small dot represents a group with their categories color-coded. We show the median of $\sigma_{\text{LOS}}$ at each $L_{19.5}$ bin with a bin width equal to $\Delta L_{19.5} = 0.4$ dex for these three types of groups as open circles (isolated CGs), hexagons (embedded CGs), and diamonds (noncompact groups), whereas the shaded areas represent the coverage of the data between the 16$^\text{th}$ and 84$^\text{th}$ percentiles respectively. For each data bin, the error of the median values are estimate by multiplying the standard error of the mean by a constant of 1.25.

As can be seen, the $\sigma_{\text{LOS}} - L_{19.5}$ relations show similar monotonic trends for isolated CGs and noncompact groups.  However, there is a systematical offset that, at given $L_{19.5}$, the compact groups show systematically larger $\sigma_{\text{LOS}}$ than the noncompact ones. This result indicates that the compactness (size) of groups might play an important role in describing the dynamics of groups of galaxies. Indeed, the galaxy groups are known to be distributed on a fundamental plane (FP) in the logarithm space of $L-\sigma-R$ parameters \citep[e.g.,][]{1998A&A...331..493A, 1999A&A...344..749F, 2005MNRAS.364.1299D, 2020A&A...641A..94D}. We will discuss the FP of isolated groups in more detail in Section~\ref{sec:dynstat-iso}.

Moreover, the median $\sigma_{\text{LOS}}$ of embedded CGs are significantly larger than that of isolated CGs at a given $L_{19.5}$. The very large offset of the embedded CGs at a given $L_{19.5}$ implies that the embedded CGs might not be a dynamically bound system. Based on the scatter of $\sigma_{\text{LOS}}$ varied with $L_{19.5}$, we suppose that the dynamical status of such embedded systems are more likely to be dependent on their parent groups because the parent groups of embedded CGs span a very large range in their dynamical mass. To verify this hypothesis, for each embedded CG, we take the total luminosity of its parent group $L_{19.5}^{\text{par}}$ \footnote{We note that the $L_{19.5}$ of the parent groups are directly taken from the sample III catalog of  Y07. As we have discussed in Section~\ref{sec:iso/emb}, the richness of the parent galaxies is significantly larger than CG themselves, the memberships of few galaxies with assigned redshifts from the nearest neighbour have negligible effects on final $L_{19.5}$. } and plot the median $\sigma_{\text{LOS}}-L_{19.5}^{\text{par}}$ relation for embedded CGs in figure~\ref{fig:L195Dyn}.  In this case, we see that the $\sigma_{\text{LOS}}-L_{19.5}^{\text{par}}$ relation is consistent with the $\sigma_{\text{LOS}}-L_{19.5}$ relation of isolated CGs. This result challenges our view of the dynamical nature of these embedded CGs. Are they distinct subsystems of larger host groups? If so, what determines their dynamical properties? Or even, is it possible that such systems are not dynamically unique in any way and are just formed or selected as a result of chance alignment? We will discuss this is in more detail in Section~\ref{sec:dynstat-emb}.

\begin{figure}
  \centering
  \subfigure{
    \includegraphics[width=1.\columnwidth]{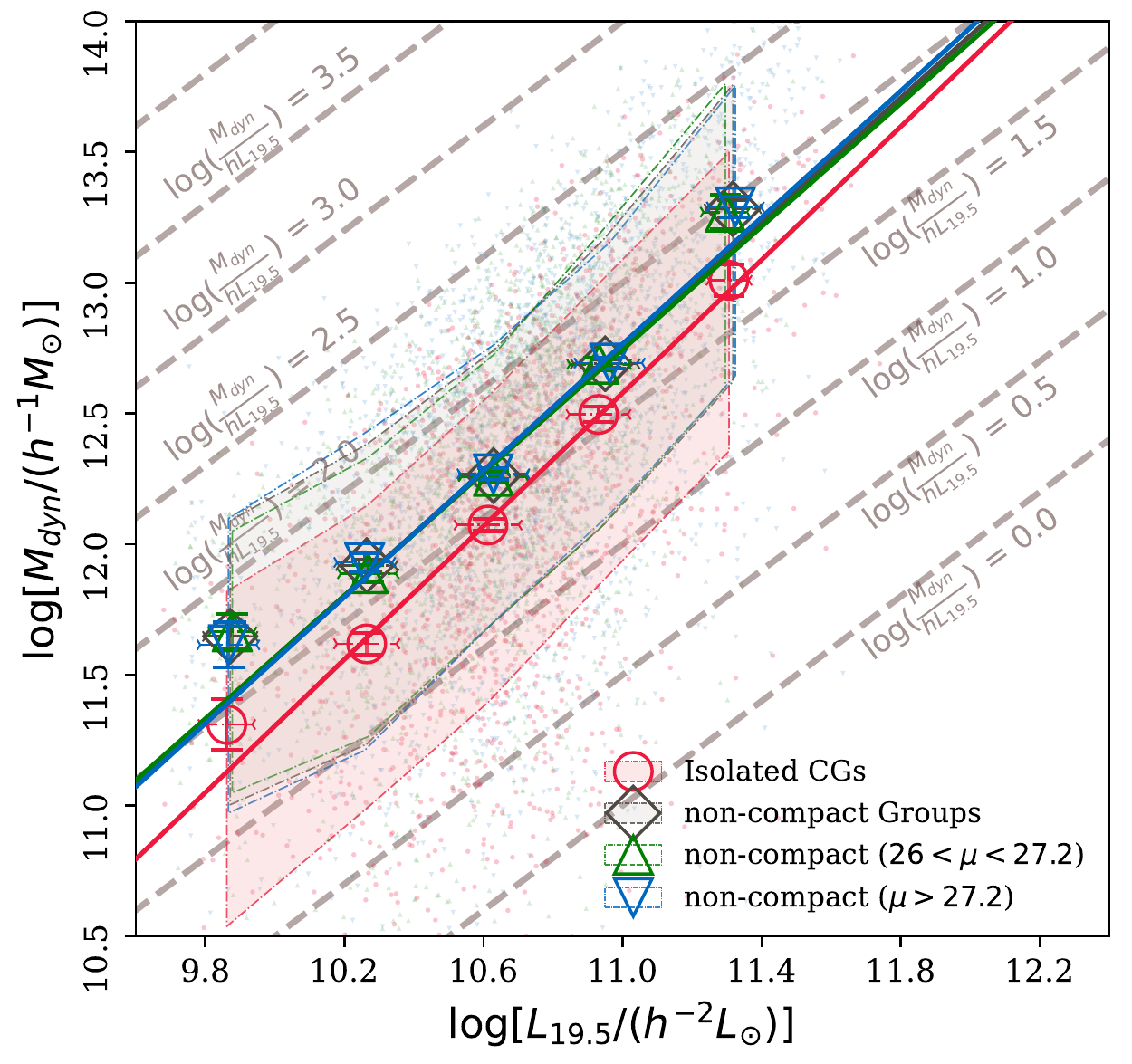}}
  \caption{Symbols and shaded areas are the median and 16$\text{th}$ to 84$\text{th}$ percentiles of $M_{\text{dyn}}$ as a function of group luminosity ($L_{19.5}$) for isolated CGs (red), and noncompact group samples (black) on a logarithmic scale with bin size of $0.4$ dex. The intermediate ($\mu \lesssim 27.2$) and loose ($\mu \gtrsim 27.2$) subsamples of noncompact groups are also shown in green and blue, respectively. The solid lines are the linear fit for these samples. Thick dashed lines are $M_\text{dyn}-L_{19.5}$ relation based on virial equilibrium ($M_{\text{dyn}} \propto L_{\text{19.5}}$) with various $\frac{M_{\text{dyn}}}{L_{\text{19.5}}}$ values. The vertical error bars show the errors of the median $\log\left(M_{\text{dyn}}\right)$ and the horizontal error bars indicate the median absolute deviation of $\log\left(L_{19.5}\right)$ in each bin. Only the data bins with at least 10 groups are plotted.}
  \label{fig:L195DynM}
\end{figure}

\subsection{Dynamical status of isolated CGs} \label{sec:dynstat-iso}
For a gravitationally bound system, $\sigma_{\text{LOS}}$ is related to its total dynamical mass in the following way: 
\begin{equation}
 M_{\text{dyn}} = \frac{{\sigma}^{2}R_{\text{dyn}}}{G}, \label{eq:dynm} \,
\end{equation}
where $\sigma^{2} = 3\sigma_{\text{LOS}}^{2}$ is 3D velocity dispersion based on isotropic assumption and $R_{\rm dyn}$ is its dynamical radius. Following \citet{1994AJ....107..868D}, we take the mean harmonic radius $R_H$ to characterize the dynamical radius given by
\begin{equation}
\frac{1}{R_{H}} = \frac{2}{N(N-1)}\sum_{i<j} \frac{1}{R_{\perp,ij}}, \label{eq:dynr}
\end{equation}
\noindent where $R_{\perp,ij}$ is the projected separation between the $i{\rm th}$ and $j{\rm th}$ members. Under the isotropic assumption, the dynamical radius should be corrected as $R_{\text{dyn}} = \frac{\pi}{2}R_{H}$ \citep{1960ApJ...132..286L}.

We plot the $M_{\text{dyn}} - L_{19.5}$ relations for isolated CGs and noncompact groups in figure~\ref{fig:L195DynM}. Moreover, we perform a linear regression (weighted by the error of the estimated $M_{\text{dyn}}$) between the median $M_{\text{dyn}}$ and $L_{\text{19.5}}$ in logarithmic space for isolated CGs and noncompact groups, respectively.  The best fits have slope $1.27 \pm 0.05$ and $1.19 \pm 0.03$ for isolated CGs and noncompact groups respectively, which are consistent with each other inside $1-\sigma$ errors. The slopes of both type groups are larger than $1$, which are qualitatively in agreement with early findings for normal galaxy groups \citep[e.g.,][]{2000ApJ...530...62G, 2005A&A...433..431P}. We argue that a slope larger than $1$ in the $M_{\text{dyn}} - L_{19.5}$ relation is a result of systematical larger mass-to-light ratio $M_{\text{dyn}}/L_{19.5}$ for higher mass groups (see the lines of constant $M_{\text{dyn}}/L_{19.5}$ ratios in figure~\ref{fig:L195DynM} for reference), which also has been suggested by early studies \citep[e.g.,][]{2000ApJ...530...62G, 2002ApJ...569..720G, 2004MNRAS.355..769E, 2007A&A...464..451P}. In this study, we are focusing on the comparison between the CGs and normal groups and therefore do not further explore the physical implications of the exact slopes of the $M_{\text{dyn}} - L_{19.5}$ relation.

For isolated CGs, comparing with the $L-\sigma$ relation shown in figure~\ref{fig:L195Dyn}, when $R_{\rm H}$ is taken into consideration, their median $M_{\text{dyn}}$ becomes systematically smaller than that of the noncompact groups. That is to say, the $R_{\rm H}$ of CGs are significantly smaller than that of noncompact groups with the same $L_ {19.5}$,  even smaller than the prediction of Equation \ref{eq:dynm}. To further identify the effect of $R_{\rm H}$ in the calculation of $M_{\rm dyn}$, we further divide the noncompact groups into two subsamples with equal numbers at their median surface brightness $\mu \sim 27.2$ mag arcsec$^{-2}$.  Here, the surface brightness of noncompact groups is calculated following the same way as that for CGs, i.e., the mean surface brightness of galaxies inside the innermost circle $\theta_G$.  Therefore, comparing with the isolated CGs $(\mu<26.0)$, one subsample of the noncompact groups is very loose ($\mu> 27.2$), and the other is intermediate ($26.0<\mu<27.2$). We show the resulting $M_{\text{dyn}} - L_{19.5}$ relations for these two subsamples as green and blue triangles in figure~\ref{fig:L195DynM} respectively. As can be seen, there are negligible differences of the $M_{\text{dyn}} - L_{19.5}$ relation between the two subsamples of noncompact groups, which implies that they might be in a quasi-virial equilibrium state so that the dynamical mass remains constant when $R_{\rm H}$ varies.

On the other hand, the systematically lower $M_{\text{dyn}}$ of isolated CGs implies that they have deviated from quasi-dynamic equilibrium and entered into a phase of galaxy merging. Specifically, when galaxy groups evolve and reach compact status of $\mu \sim 26$ mag arcsec$^{-2}$, the internal frequent close encounters of galaxies cause dynamical friction and shrink the mean separation between the group members significantly, which therefore results in a smaller $M_{\text{dyn}}$ being measured. We present a more detailed discussion on the dynamical evolution of galaxy groups in Appendix~\ref{sec:scheme}.

\begin{figure*}
  \centering
  \subfigure{
    \includegraphics[width=2.\columnwidth]{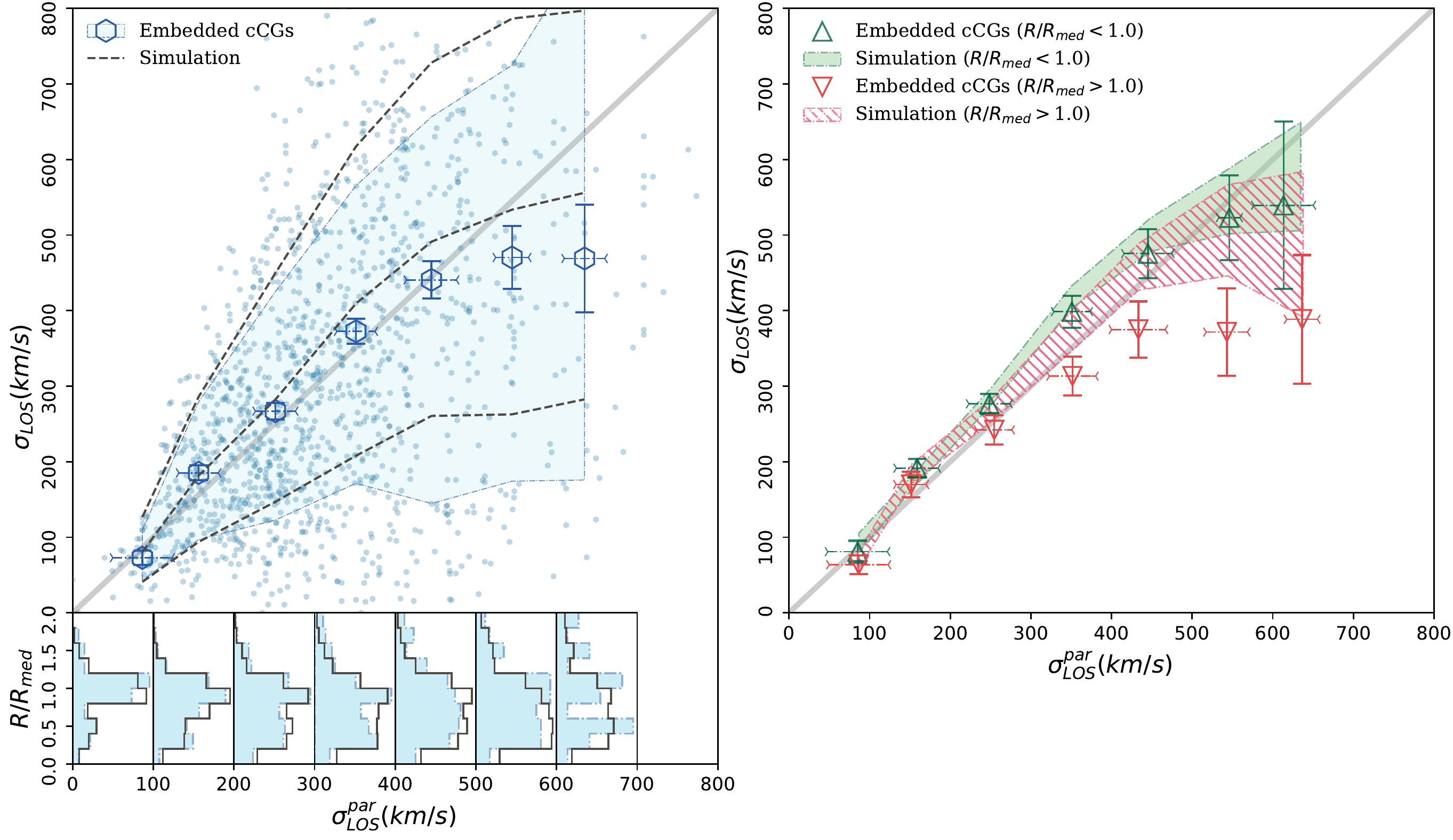}}
  \caption{The comparison of the embedded CGs and mock CGs from Monte Carlo simulation. Left: velocity dispersion of the embedded CGs ($\sigma_{\text{LOS}}$) versus that of their parent groups ($\sigma_{\text{LOS}}^{\text{par}}$). The open hexagons represent the median $\sigma_{\text{LOS}}$ of the embedded CGs in each $\sigma_{\text{LOS}}^{par}$ bin, whose vertical error bars represent the errors of the median $\sigma_{\text{LOS}}$ and horizontal error bars represent the median absolute deviation of $\sigma_{\text{LOS}}^{\text{par}}$ in each bin. The shaded area shows the 16$\text{th}$ and 84$\text{th}$ percentiles of the $\sigma_{\text{LOS}}$ distribution at a given $\sigma_{\text{LOS}}^{\text{par}}$, whereas the three dashed lines show the 16$\text{th}$, 50$\text{th}$, and 84$\text{th}$ percentiles for the mock CGs.  The bottom shows the histograms of the radial distance of observed (filled) and mock (open) CGs within the parent groups in each $\sigma_{\text{LOS}}^{\text{par}}$ bin. Right: The CGs located at $R/R_{\text{med}}>1$ (red) and $R/R_{\text{med}}<1$ (green), respectively. The triangles show the observed embedded CGs, while the shaded and hatched areas represent the standard deviation of the median $\sigma_{\text{LOS}}$ of mock CGs, respectively. The bold lines in both panels are the one-to-one correspondence between $\sigma_{\text{LOS}}$ and $\sigma_{\text{LOS}}^{\text{par}}$.}
  \label{fig:MhDyn}
\end{figure*}

\subsection{Dynamical nature of embedded CGs} \label{sec:dynstat-emb}
To have a better understanding of the dynamical nature of the embedded CGs, we directly compare the relation between $\sigma_{\text{LOS}}$ and $\sigma_{\text{LOS}}^{\text{par}}$ for embedded CGs and show the results in figure~\ref{fig:MhDyn}. 

There is a good one-to-one correlation for these groups with $\sigma_{\text{LOS}}^{\text{par}} \lesssim 500$ km s$^{-1}$.  As we have mentioned in Section~\ref{sec:iso/emb}, all parent groups have total luminosities at least twice of their embedded CGs (Equation~\ref{eq:emb}) and their typical richness are $N \gtrsim 8$ (Figure~\ref{fig:extra} of Appendix~\ref{sec:parg}). Therefore, we do not expect that the good consistency between $\sigma_{\text{LOS}}$ and $\sigma_{\text{LOS}}^{\text{par}}$ is caused by the dominance of the embedded CG members in parent groups.  The excellent consistency between $\sigma_{\text{LOS}}$ and $\sigma_{\text{LOS}}^{\text{par}}$ implies that the dynamics of these apparent subsystems are only determined by their host groups. That is to say, these embedded CGs are not dynamically distinct subsystems but more likely to be the consequence of chance alignments within larger systems. We will verify this hypothesis using a detailed Monte Carlo simulation below.

For massive clusters of galaxies ($\sigma_{\text{LOS}}^{\text{par}} \gtrsim 500$ km s$^{-1}$), the median $\sigma_{\text{LOS}}$  are on average $20\% - 40 \%$ below the one-to-one prediction. This might be caused by the bias of the velocity difference criterion ($|V - V_{\text{med}}| < 1000$ km s$^{-1}$) used in the CG selection, which the galaxies with large relative velocities chosen randomly from large $\sigma_{\text{LOS}}^{\text{par}}$ groups are more likely to violate. By excluding such systems from CG selection, the resulting embedded CGs would certainly have $\sigma_{\text{LOS}} < \sigma_{\text{LOS}}^{\text{par}}$ on average. On the other hand, it is also possible that these embedded CGs are gravitationally bound subsystems that have recently fallen into large clusters.  In this case, the dynamics of their members could be partially heated but still be kinematically colder than the host clusters \citep[e.g.,][]{2019MNRAS.490.3654C, 2020MNRAS.tmp.2076B}). To distinguish these two different scenarios, we also need a Monte Carlo simulation. \\

\noindent \textit{Monte Carlo Simulation:} We run a Monte Carlo simulation to test the hypothesis that the embedded CGs are purely selected from the chance aliment of the galaxy members in parent groups.  To make the simulation as realistic as possible, we build the mock CGs from the parent groups of embedded CGs. Specifically, for each parent group, we keep the projected radial distances of all members with respect to the luminosity-weighted center defined by Y07 and keep their radial velocities unchanged, only randomizing their projected azimuthal positions. We then apply the CG selection criteria used in Paper I and the embedded criterion (Equation~\ref{eq:emb}) to search the mock CGs. For each embedded CG, we perform multiple runs by randomizing its parent group until 100 mock CGs were derived ($100 \times 1370$ mock CGs overall). 

We show the median and 16$^\text{th}$ and 84$^\text{th}$ percentiles of $\sigma_{\text{LOS}}$ of the mock CGs as black dashed lines in the left panel of figure~\ref{fig:MhDyn}. In addition to $\sigma_{\text{LOS}}$, the projected radial position distribution of the mock CGs can also be used to test the chance alignment hypothesis because the radial positions of all group members have been retained during simulation. Here we use $R/R_{\text{med}}$ to characterize the projected radial position of the embedded CGs in parent groups, where $R$ is the projected distance of the CG luminosity-weighted center to the parent group center and $R_{\text{med}}$ is the median projected distance of all members of the parent groups. The $R/R_{\text{med}}$ distributions for both of the embedded CGs and mock CGs in each $\sigma_{\text{LOS}}^{\text{par}}$ bin are also shown in the left panel of figure~\ref{fig:MhDyn}. 

As can be seen, both $\sigma_{\text{LOS}}$ and $R/R_{\text{med}}$ distributions of the embedded CGs with $\sigma_{\text{LOS}}^{\text{par}} \lesssim 300$ km s$^{-1}$ can be well reproduced by the mock CGs, which validates the chance alignment hypothesis. However, it is worth mentioning that although the embedded CGs have been selected in a way that they do not dominate the luminosity of their hosts, the richness of their parent groups are not necessarily at least 2 times higher. Few extra bright galaxies in the parent group would be sufficiently double the luminosity. In this case, the $\sigma_{\text{LOS}}^{\text{par}}$ would be dominated by the $\sigma_{\text{LOS}}$ of embedded CGs, and we could not distinguish a dynamically bound CG from the chance alignment hypothesis.

There is a slight deviation between the observed and mock CGs within the $300$ km s$^{-1}$ $< \sigma_{\text{LOS}}^{\text{par}} < 500$ km s$^{-1}$ groups and tend to be very significant for $\sigma_{\text{LOS}}^{\text{par}} > 500$ km s$^{-1}$ groups. For the groups with $\sigma_{\text{LOS}}^{\text{par}} > 500$ km s$^{-1}$, although mock CGs have already passed the velocity filter ($V-V_{\text{med}} > 1000$ km s$^{-1}$), there still exists the systematic deviation from observation. Moreover, the predicted $R/R_{\text{med}}$ distribution also shows significant differences from the observations. The embedded CGs we identified in $\sigma_{\text{LOS}}^{\text{par}} > 300$ km s$^{-1}$ groups are evidently and gradually biased to the outer regions of the parent groups. Combing these two effects, we infer that, for the large-scale environment like group systems with $\sigma_{\text{LOS}}^{\text{par}} > 300$ km s$^{-1}$, the CGs we selected using traditional Hickson-like criteria might not be fully explained by the chance alignment effect. Some of these compact subsystems might be bound objects that could have entered into the larger systems in the last $1 - 2$ Gyr as argued by \citet{2018ApJ...865...40L}. For these infalling systems, they might have yet to complete their first pericentric passage due to the short time-scale, which makes them mainly be located at the outer regions of the parent system and with their morphology being kept in a compact state.

To further verify this conclusion, we separate embedded CGs into two categories according to their radial position and compare their $\sigma_{\text{LOS}}$. We show the median $\sigma_{\text{LOS}}$ of the inner ($R/R_{\text{med}} < 1$) and outer ($R/R_{\text{med}} > 1$) CGs as the green and red triangles in the right panel of figure~\ref{fig:MhDyn}, the standard deviation of the median $\sigma_{\text{LOS}}$ of corresponding mock CGs obtained from 1000 bootstrap resamplings with the same sample size as observed CGs in each bin are shown by the green filled and red dashed areas, respectively. Clearly, the $\sigma_{\text{LOS}}$ of inner CGs can be well reproduced by Monte Carlo simulation. This result implies that the inner CGs might be dominated by chance alignments along the LOS within their parent groups \citep{1986ApJ...307..426M, 2008A&A...486..113M}, where the high number density of galaxy members in central regions of galaxy groups could easily trigger such a selection bias. Conversely, the $\sigma_{\text{LOS}}$ of the outer CGs are systematically below the mock samples. This result indicates that these outer CGs, at least, might consist of (or include) newly accreted groups. Indeed, numerical simulations \citep[e.g.,][]{2012MNRAS.419.1017C, 2019MNRAS.490.3654C} have predicted that these newly accreted systems have not passed through the cluster center and are mainly located at the outskirts of host clusters.  Despite of experiencing dynamical heating, these subsystems remain dynamically colder and more compact than the host cluster \citep{2020MNRAS.tmp.2076B}.  After the first pericentric passage, they would soon be disassembled and virialized within the host cluster. Such a scenario is also consistent with the early finding that the substructures of rich clusters appear to decrease toward their central regions \citep{2002A&A...387....8B}.

\begin{figure*}
  \centering
  \subfigure{
    \includegraphics[width=1.\columnwidth]{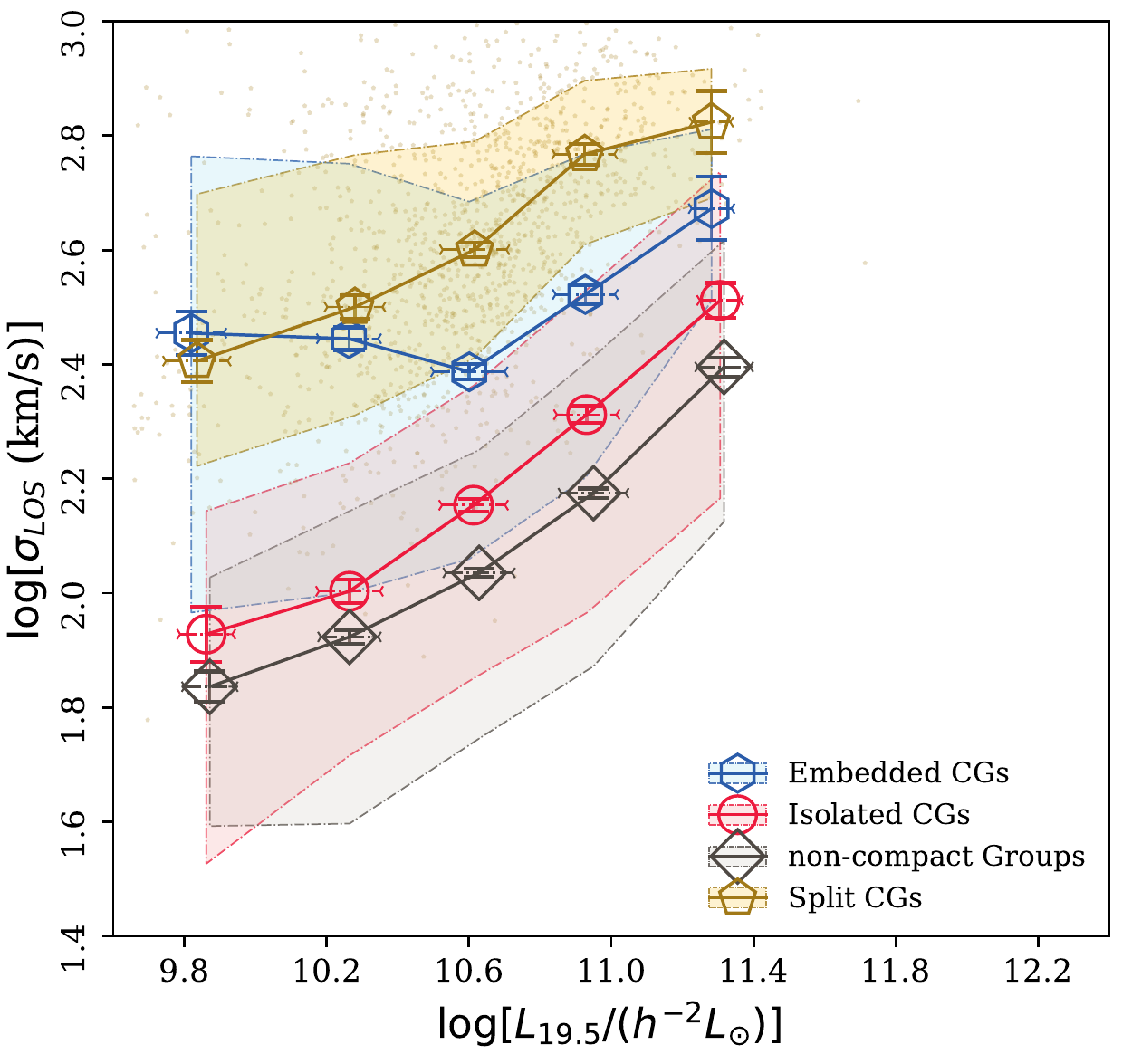}}
  \subfigure{
    \includegraphics[width=1.\columnwidth]{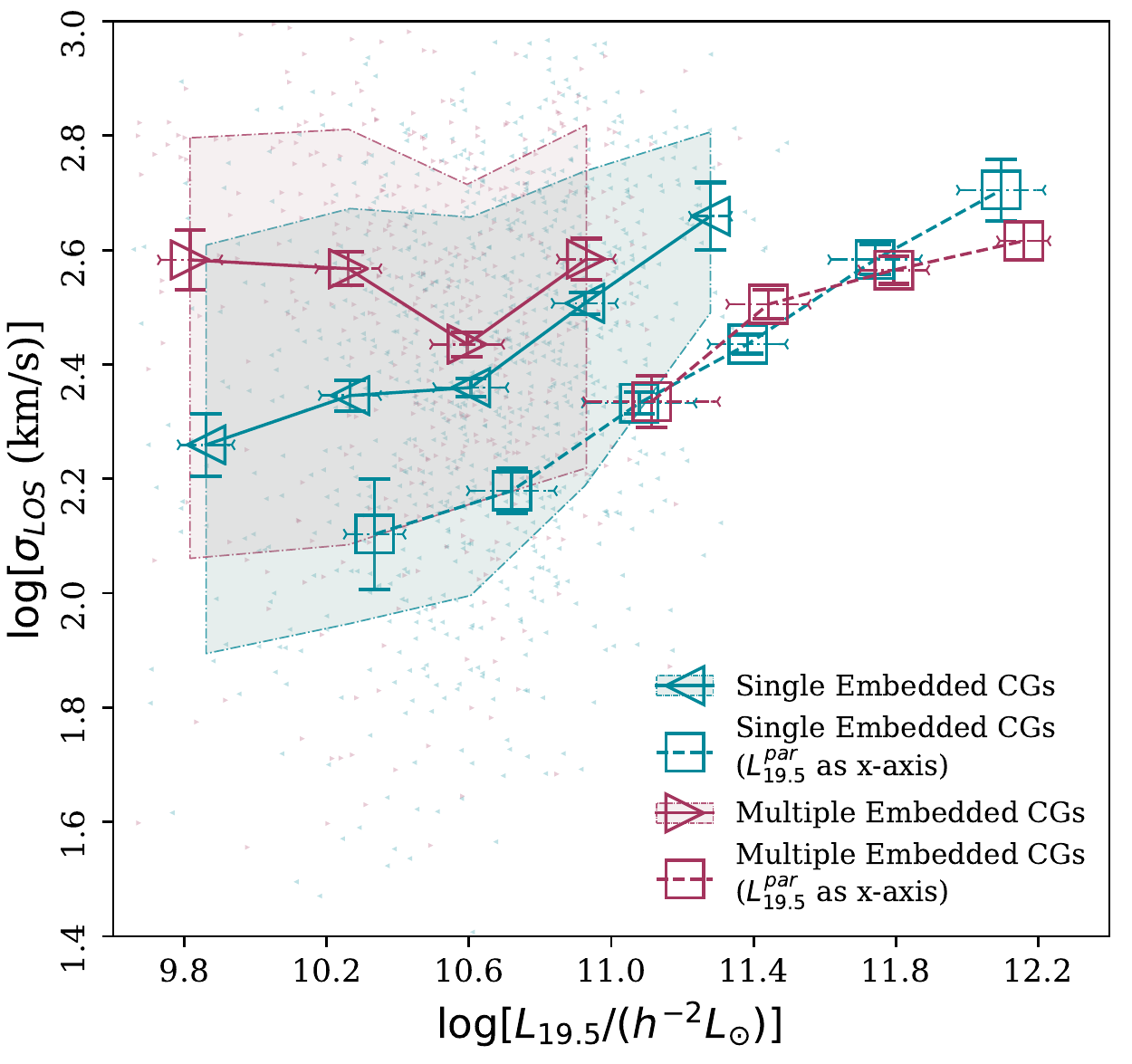}}
  \caption{The $L_{19.5}-\sigma_{\rm LOS}$ relation for the subtypes of CGs that we have not discussed in detail in the main text. Left panel: split CGs (gold), compared with noncompact groups (black), isolated (red), and embedded CGs (blue). Right panel: single (teal) and multiple (purple) embedded CGs, where the symbols are connected by dashed lines that represent the same scale relation but use $L_{19.5}^{\text{par}}$ as the x-axis. The vertical error bars show the errors of the median $\log\left(M_{\text{dyn}}\right)$ and the horizontal error bars indicate the median absolute deviation of $\log\left(L_{19.5}\right)$ in each bin. Only the data bins with at least 10 groups are plotted.}
  \label{fig:multi}
\end{figure*}

\begin{figure*}
  \centering
  \subfigure{
    \includegraphics[width=.8\columnwidth]{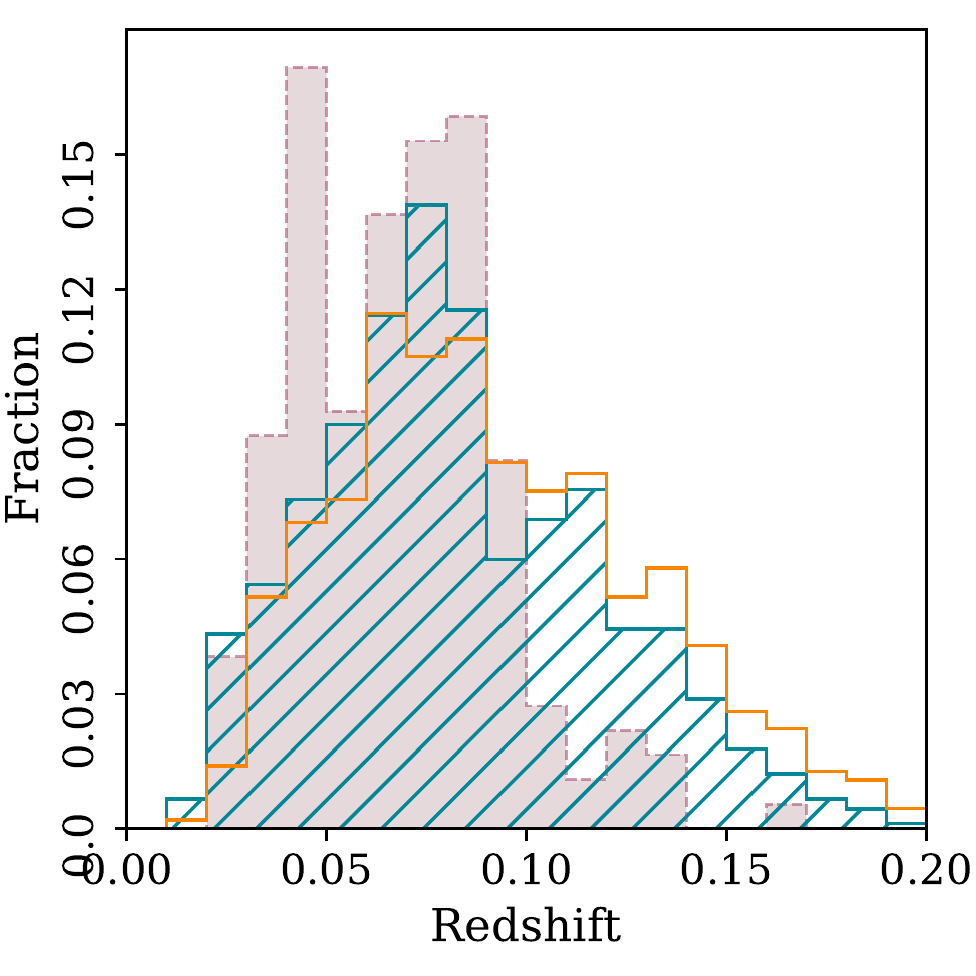}}
  \subfigure{
    \includegraphics[width=.8\columnwidth]{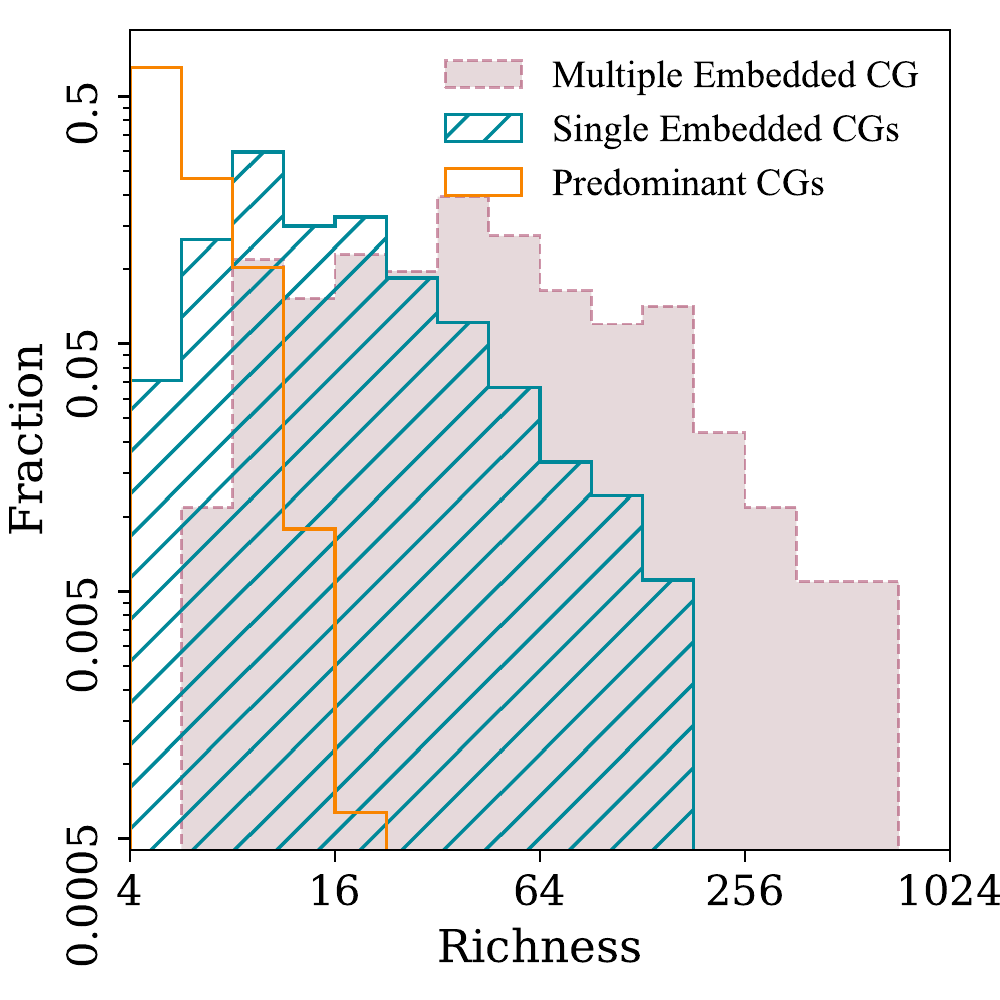}}
  \subfigure{
    \includegraphics[width=.8\columnwidth]{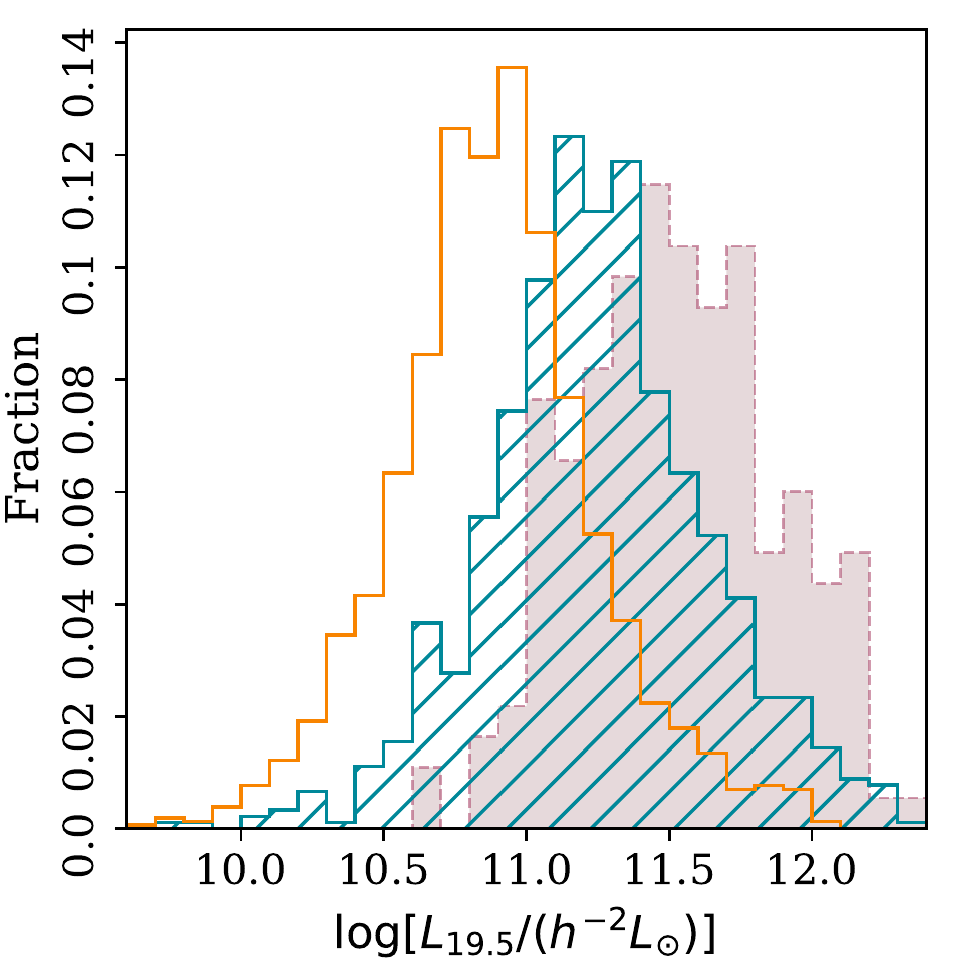}}
  \subfigure{
    \includegraphics[width=.8\columnwidth]{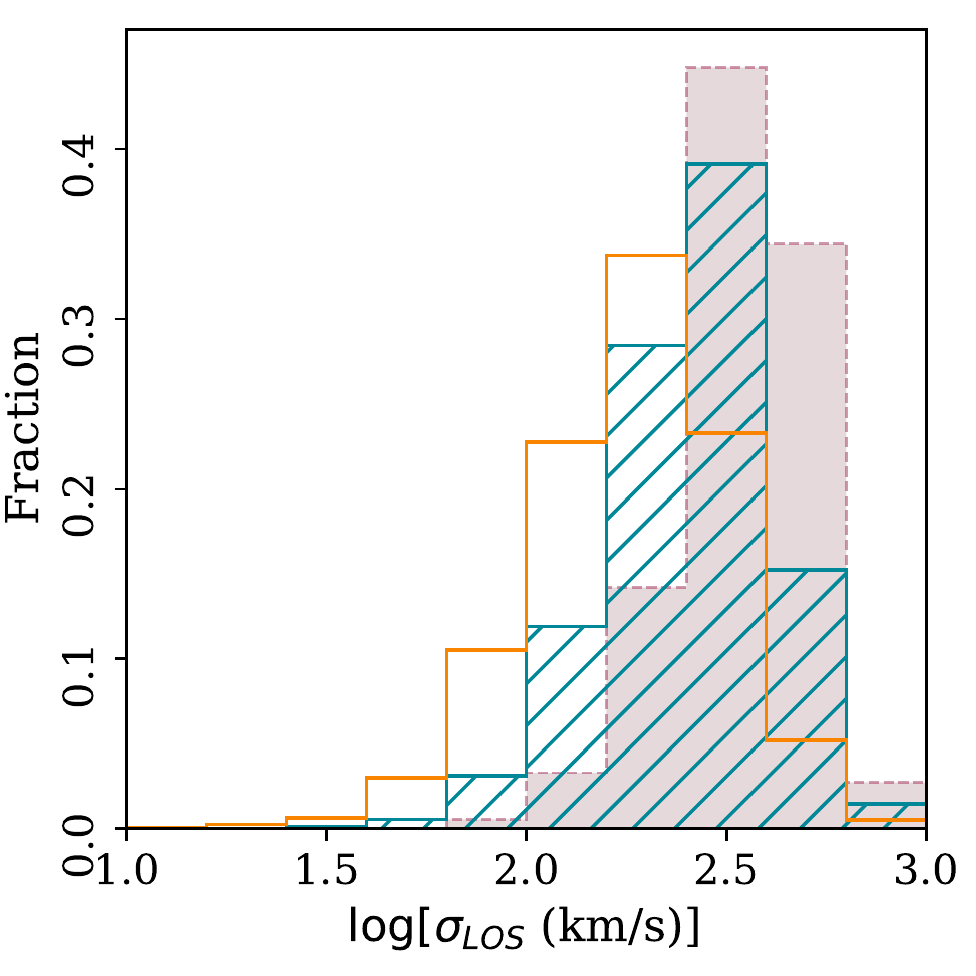}}
  \caption{The distributions of the parent groups of predominant (open), single embedded (hatched), and multiple (filled) embedded CGs. Upper left: redshift. Upper right: richness. Lower left: $L_{19.5}$. Lower right: LOS velocity dispersion. }
  \label{fig:extra}
\end{figure*}

\section{Conclusions} \label{sec:summarize}
In this paper, we use a large sample of CGs taken from Paper I to explore their spatial relation with Y07 groups defined by the halo-based model. We show that $\sim 27 \%$ of CGs have a one-to-one correspondence with the groups in a single dark matter halo, which we refer to as ``isolated CGs". The remaining CGs have complex associations with dark matter halos. After removing the CGs that dominate the luminosity of the haloes ($\sim 26 \%$) and linked to multiple halos ($\sim 21 \%$), there are $\sim 23 \%$ of them embedded within large clusters as nondominant components, which we refer to as ``embedded CGs." The relatively low percentage of isolated CGs we found is a result of our careful inspection of the relations between CGs and the halo-based group sample. In our result, the fraction ratio of isolated CGs to embedded CGs is about $1:1$, which is consistent with the early finding of \citet{2011MNRAS.418.1409M}. If we consider these CGs that dominate the luminosity of host halos also to be isolated CGs also, we will get a ratio of $2.3:1$, which then is in good agreement with the result of  \citet{2015A&A...578A..61D}.

We take advantage of the velocity dispersion measurements of our CG sample and use it to explore the dynamical features of isolated and embedded CGs, respectively. Our main results are as follows:
\begin{enumerate}
\setlength{\itemsep}{-0.7ex}
\item The correlation between velocity dispersion ($\sigma_{\text{LOS}}$) and group luminosity ($L_{19.5}$) for isolated CGs is monotonic and similar to that for noncompact groups. However, the $\sigma_{\text{LOS}}$ of isolated CGs are systematically higher than that of noncompact groups at a given $L_{19.5}$. By considering the group radius, we find that the dynamical mass of isolated CGs is systematically smaller than that of noncompact groups. But for noncompact groups, their dynamical mass shows negligible dependence on group compactness. This result implies that the isolated CGs are more likely to be dynamically more evolved systems that have entered into the orbital dissipation phase induced by dynamical friction.
\item For embedded CGs, the correlation between their $\sigma_{\text{LOS}}$ and $L_{19.5}$ of themselves is much weaker than with the $L_{19.5}$ of their parent groups ($L_{19.5}^{\text{par}}$).  This result indicates that the dynamical status of embedded CGs might depend on their parent groups. Using a more detailed Monte Carlo simulation, we further show that the embedded CGs located at the inner regions of dark matter halos are more likely to be caused by the chance alignment effect of high-density regions. On the other hand, the CGs embedded at the outer regions of dark matter halos might contain a certain fraction of infalling groups accreted in the last 1 - 2 Gyr.
\end{enumerate}

Our dynamical analysis has revealed that the observationally selected CGs are a heterogeneous system.  More detailed studies are required to further reveal their physical natures. For CGs embedded in rich groups or clusters, we need a better diagnostic tool to distinguish the infalling groups from the chance alignment effect. To do that, a deeper spectroscopy survey (e.g., DESI \citealp{2016arXiv161100036D}) could significantly reduce the random fluctuations by increasing the number of the member galaxies. On the other hand, the physical properties of the member galaxies  are helpful for resolving the connections between the embedded CGs and their host environment. For isolated CGs, their distinct dynamical features imply that the compactness of galaxy groups might be an important indicator describing the physical properties of galaxy groups. A more detailed study of the correlation between the compactness of galaxy groups and the physical properties of their member galaxies is expected in an upcoming study, which will help us to have a better understanding of the formation and evolution of galaxy groups.

\acknowledgments
We gratefully acknowledge the anonymous referee for helpful comments and detailed suggestions. This work is supported by the National Key R\&D Program of China (No. 2019YFA0405501), the National Natural Science Foundation of China (No.12073059), and the Cultivation Project for LAMOST Scientific Payoff and Research Achievement of CAMS-CAS.

This work has made use of data products from the Sloan Digital Sky Survey (SDSS, \url{http://www.sdss.org}), the Large Sky Area Multi-Object Fiber Spectroscopic Telescope (LAMOST, \url{http://www.lamost.org}), the GAMA survey (\url{http://www.gama-survey.org}. We are thankful for their tremendous efforts in the surveying work.

The Guoshoujing Telescope (the Large Sky Area Multi-Object Fiber Spectroscopic Telescope LAMOST) is a National Major Scientific Project built by the Chinese Academy of Sciences. Funding for the project has been provided by the National Development and Reform Commission. LAMOST is operated and managed by the National Astronomical Observatories, Chinese Academy of Sciences.

\begin{appendix}
\section{The $\sigma_{\text{LOS}} - L_{19.5}$ Relation for Split, Single Embedded and Multiple Embedded CGs} \label{sec:multi}
In this appendix, we show the $\sigma_{\text{LOS}} - L_{19.5}$ relations for two subtypes of CGs that we have not discussed in detail in the main text of the manuscript, the split CGs and multiple embedded CGs. The split CGs are those that inhabit multiple Y07 groups (an example is shown in the panel (d) of figure~\ref{fig:CaseN}), whereas the multiple embedded CGs are the cases where at least two CGs embedded in the same Y07 group.  For split CGs, their $L_{19.5}$ are corrected by the LF of the isolated CGs for simplicity.  The results are shown in the left and right panels of figure~\ref{fig:multi} respectively.

At a given group luminosity $L_{19.5}$, the split cCGs show significantly higher $\sigma_{\text{LOS}}$ than isolated CGs and even single embedded CGs. As we have already mentioned in the main text, the high velocity dispersion of split CGs are mainly attributed to the large velocity difference cut $\Delta V < 1000$  km s$^{-1}$ used in CG selection (Paper I). Therefore, we conclude that these split CGs are apparent systems and are not gravitationally bound. 

In the right panel of figure~\ref{fig:multi}, we compare $\sigma_{\text{LOS}} - L_{19.5}$ and  $\sigma_{\text{LOS}} - L_{19.5}^{\text{par}}$ for single embedded and multiple embedded CGs. Apparently, at a given $L_{19.5}$ of CGs, the multiple embedded CGs show higher $\sigma_{\text{LOS}}$ than a single embedded one. When considering the luminosity of parent groups $L_{19.5}^{\text{par}}$, the multiple and single embedded CGs follow the same $L-\sigma$ relation. Therefore, the $\sigma_{\rm LOS}$ of both single and multiple embedded CGs are dominated by their parent groups, and the higher $\sigma_{\rm LOS}$ of multiple embedded CGs is simply a result of their richer parent groups.

\section{The basic statistical properties of Parent Groups} \label{sec:parg}
In figure~\ref{fig:Basic}, we have presented the basic statistical properties of different categories of CGs when matched with Y07 groups. Here we show the statistical properties of the Y07 groups that host our CG samples.

As mentioned in section~\ref{sec:iso/emb}, apart from 1667 isolated CGs and 1282 split CGs, there are 1570 and 1370 predominant and embedded CGs respectively, the latter of which include 469 multiple embedded CGs and 901 single embedded CGs. Figure~\ref{fig:extra} shows the histograms of redshift, richness, $L_{19.5}$, and $\sigma_{\text{LOS}}$ for the parent groups of the predominant and both types of embedded CGs. 

These three categories of parent groups show significantly different redshift and richness distributions, which, however, are obviously results of selection effects. From predominant CGs, single embedded CGs, to multiple embedded CGs, because of their decreasing dominance and similar richness distribution, their parent groups certainly have increasing richness (upper-right panel) and  $L_{19.5}$ (lower-left panel) distributions. Their different $\sigma_{\text{LOS}}$ distributions (lower-right panel) are then simply the result of the $\sigma_{\text{LOS}} - L_{19.5}$ relation. For the redshift distribution, because of the selection effects in a flux-limited sample, the higher richness parent groups are certainly biased to lower redshifts (upper-left panel).

\begin{figure}
  \centering
  \subfigure{
    \includegraphics[width=1.\columnwidth]{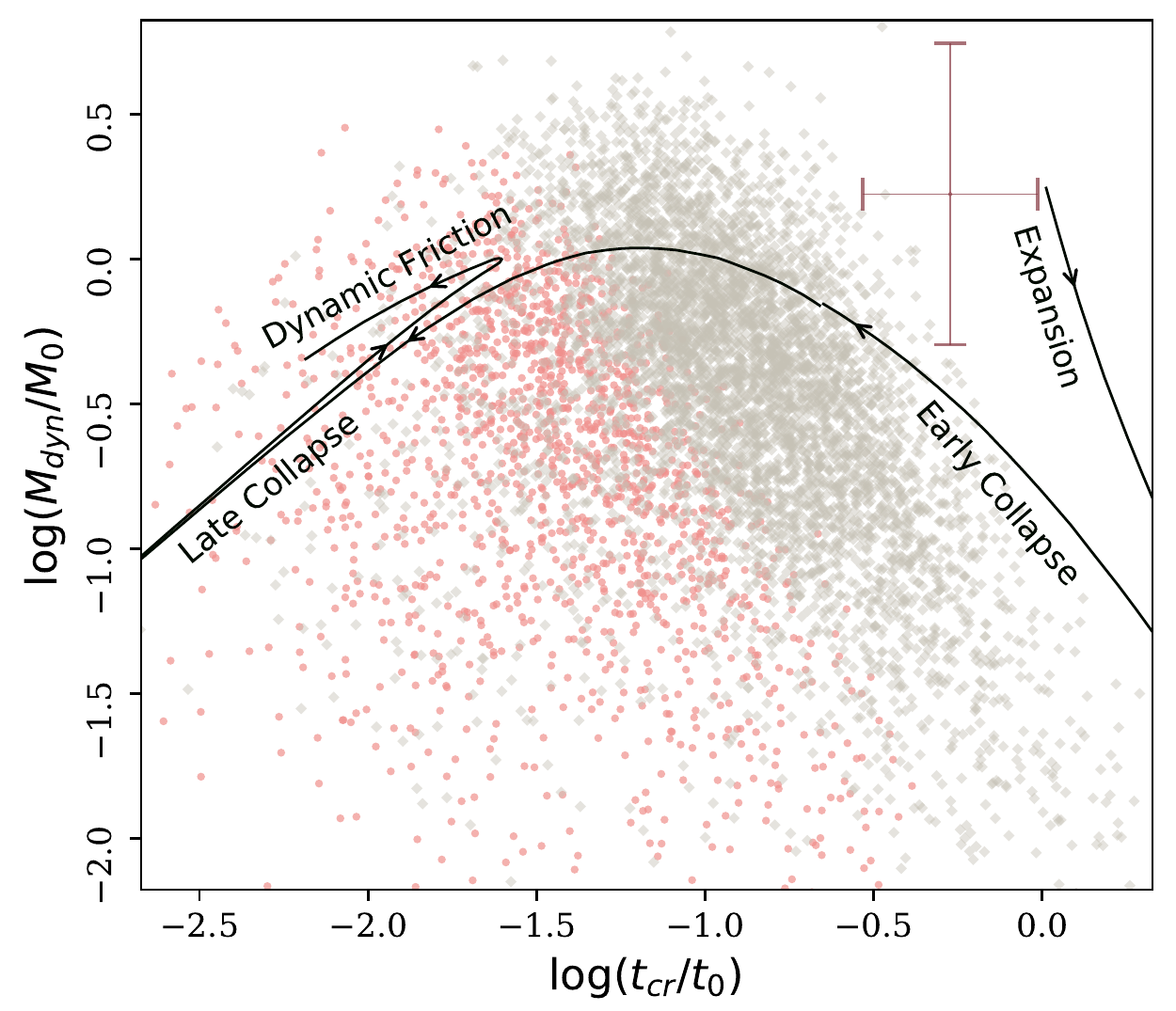}}
  \caption{The dynamic mass of groups scaled to their true mass ($\frac{M_{\text{dyn}}}{M_{0}}$) versus crossing time ($t_{\text{cr}}$) scaled to the age of universe $t_0$ for isolated CGs (red) and noncompact groups (gray), respectively. The error bar represents the typical error of these samples. The solid lines are the evolutionary track of a galaxy system adapted from \citet{1993nbpg.conf..188M} based on the softened potential assumption and the arrows indicate the evolutionary direction.}
  \label{fig:M94}
\end{figure}

\section{The evolutionary scheme for groups} \label{sec:scheme}

\citet{1993nbpg.conf..188M, 2007ggnu.conf..203M} has provided an analytical model of the dimensionless mass bias, $M_{\text{dyn}}/M_{0}$, where $M_{0}$ refers to the true mass of the galaxy systems, versus the dimensionless crossing time $t_{\text{cr}}/t_{0}$, where $t_{0}$ refers to the age of the universe, for an isolated system at different evolutionary stage. Figure~\ref{fig:M94} shows the solid track that galaxy systems should follow, the arrows indicate the evolutionary direction of a galaxy system: expands alongside with the Hubble flow at first, then decouples from this flow, turns around when reaches maximum expansion, collapses subsequently, and finally virializes with continuous orbital energy dissipation induced by dynamical friction. 

In this appendix, we plot the $M_{\text{dyn}}/M_0$ versus $t_{\text{cr}}/t_0$  relation for isolated CGs and noncompact groups so as to further compare their dynamical status.  The isolated CGs and noncompact groups are plotted as red and gray dots, respectively, in figure~\ref{fig:M94} where their crossing time is given by $t_{\text{cr}} = R_{\text{dyn}}/\sigma$. Unfortunately, true masses of galaxy systems, $M_{0}$, are unknown. Here, we have made the simple assumption of $M_{0}/(hL_{19.5}) \sim 150$ \citep[see][]{2007ApJ...671..153Y}. The majority of galaxy groups are located near the theoretical track within the error tolerance, and the isolated CGs are generally in  more evolved region compared with the noncompact ones. Considering possible shift of y-axis (due to unknown $M_{0}$), figure~\ref{fig:M94} shows that many of the isolated CGs are more likely to be in the phase of dynamic friction.

\end{appendix}

\newpage
\bibliographystyle{aasjournal}
\bibliography{ms2m.bib}

\makeatother
\end{CJK*}
\end{document}